\documentclass[11pt]{article}
\pdfoutput=1
\usepackage[pdftex]{hyperref}
\hypersetup{
    colorlinks,%
    citecolor=blue,%
    filecolor=black,%
    linkcolor=blue,%
    urlcolor=blue
}
\usepackage{amsfonts}
\usepackage{amsmath}
\usepackage{latexsym}
\usepackage{bbm}
\usepackage[T1]{fontenc}
\usepackage[latin1]{inputenc}
\usepackage{geometry}
\usepackage{booktabs}
\usepackage{amsthm}
\usepackage{color}
\usepackage{epsfig}
\usepackage{psfrag}
\usepackage{lscape}
\usepackage{graphicx}
\usepackage{appendix}

\newcommand{\CP}{CP^{3}}

\newcommand{\beq}{\begin{equation}}
\newcommand{\eeq}{\end{equation}}
\newcommand\beqa{\begin{eqnarray}}
\newcommand\eeqa{\end{eqnarray}}
\newcommand\bea{\begin{array}}
\newcommand\eea{\end{array}}
\newcommand{\COMMENT}[1]{{}}

\newcommand{\neqa}{\nonumber\end{eqnarray}}

\def\and{{\rm and}}

\begin{document}
\vspace*{-.6in} \thispagestyle{empty}
\begin{flushright}
CALT 68-2760
\end{flushright}
\baselineskip = 18pt

\vspace{0.4in}

\begin{center}
{\Large \textbf{One-Loop Corrections to Type IIA String Theory in $%
AdS_{4}\times CP^{3}$}}
\end{center}


\begin{center}
Miguel A. Bandres and Arthur E. Lipstein \\[0pt]
\emph{California Institute of Technology\\[0pt]
Pasadena, CA 91125, USA} \\[0pt]
\emph{bandres@caltech.edu, arthur@theory.caltech.edu}
\end{center}

\vspace{0.3in}

\begin{center}
\textbf{Abstract}
\end{center}

\begin{quotation}
\noindent We study various methods for computing the
one-loop correction to the energy of classical solutions to type IIA string
theory in $AdS_{4}\times CP^{3}$. This involves computing the spectrum of
fluctuations and then adding up the fluctuation frequencies. We focus on two classical solutions with support in $CP^{3}$: a
rotating point-particle and a circular spinning string with two angular
momenta equal to $J$. For each of these solutions, we compute the spectrum of fluctuations using two techniques, known as the
algebraic curve approach and the world-sheet approach. If we use the same prescription for adding fluctuation frequencies that was used for type IIB string theory in $AdS_{5}\times S^{5}$, then we find that the world-sheet spectrum gives convergent one-loop corrections but the algebraic curve spectrum gives divergent ones. On the other hand, we find a new
summation prescription which gives finite results when applied to both the algebraic curve and world-sheet spectra.
Naively, this gives three predictions for the one-loop correction to the spinning string energy (one from the algebraic curve and two from the world-sheet), however we find that in the large-$\mathcal{J}$ limit (where $\mathcal{J}={J}/\sqrt{2 \pi^2 \lambda}$), the $\mathcal{J}^{-2n}$ terms in all three cases agree. We therefore obtain a unique prediction for the one-loop correction to the spinning string energy.

\end{quotation}

\newpage

\pagenumbering{arabic}

\tableofcontents

\section{Introduction}

One of the most intriguing ideas in theoretical physics is the $AdS/CFT$
correspondence, which relates string theory or M-theory on a background
geometry consisting of $AdS_{d+1}$ times some compact space to a
conformal field theory living in $d$-dimensional Minkowski space.
By far the best studied example of this correspondence is the duality
between type IIB superstring theory on $AdS_{5}\times S^{5}$ and $\mathcal{N}=4$
SYM proposed by Maldacena in~\cite{Maldacena:1997re}. A little over a year ago, Aharony, Bergman, Jafferis, and Maldacena
(ABJM) discovered a new example of this duality which relates type IIA string
theory on $AdS_{4}\times \CP$ to a three dimensional $\mathcal{N}=6$
Chern-Simons theory \cite{Aharony:2008ug}. Some basic properties of this gauge theory are reviewed in Appendix A.

Since then, much of the analysis that was done
to test the $AdS_{5}/CFT_{4}$ duality has been repeated for the $AdS_{4}/CFT_{3}$
duality. For example, various sectors of the planar Chern-Simons
theory were shown to be integrable up to four loops in perturbation
theory, i.e. it was shown that the dilatation operator in these sectors corresponds
to a spin-chain Hamiltonian which can be diagonalized by solving Bethe
equations \cite{Minahan:2008hf,Zwiebel:2009vb,Minahan:2009te,Bak:2009mq}. Moreover, the classical string theory dual to the planar
gauge theory was also shown to be integrable, i.e. the equations of
motion for the string theory sigma model were recast as a flatness
condition for a certain one-form known as the Lax connection \cite{Bena:2003wd,Stefanski:2008ik, Arutyunov:2008if,D'Auria:2008cw}.
It should be noted that classical integrability has
only been demonstrated in the subsector of
the $AdS_{4}\times \CP$
superspace described by the $OSp(6|4)/(U(3)\times SO(3,1))$ supercoset, and that $\kappa$-symmetry in the coset sigma model breaks down for string solutions that move purely in $AdS_{4}$ \cite{Arutyunov:2008if}. Demonstrating integrability in the full superspace requires more general methods \cite{Gomis:2008jt}. The pure spinor string theory on $AdS_4 \times \CP$ was studied in \cite{Fre:2008qc,Bonelli:2008us}. An important
consequence of the Lax connection is that any classical solution to
the sigma model equations of motion can be mapped into a multi-sheeted
Riemann surface known as an algebraic curve \cite{SchaferNameki:2004ik,Kazakov:2004qf,Beisert:2005bm}. The $AdS_{4}/CFT_{3}$
algebraic curve was constructed in~\cite{Gromov:2008bz}. Following these developments,
a set of all-loop Bethe equations, which interpolate between the gauge
theory Bethe equations at weak coupling and the string theory algebraic
curve at strong coupling, were proposed in~\cite{AllLoop}. The all-loop Bethe ansatz is a powerful tool for testing the $AdS/CFT$ correspondence.

While the $AdS_{4}/CFT_{3}$ duality shares certain features with the $AdS_{5}/CFT_{4}$ duality, it also exhibits several new
features. For example, when one looks at quantum excitations to the string
theory sigma model in the Penrose limit of type IIA string theory on $AdS_{4}\times \CP$,
one finds that half of the excitations are twice as massive as the other half \cite{Nishioka:2008gz, Grignani:2008is, Gaiotto:2008cg}. The latter are subsequently referred to as {}``light'' and the former are referred to as {}``heavy''. This is in contrast to what was found when looking at the Penrose limit of type IIB string theory on $AdS_{5}\times S^{5}$, where all the
excitations have the same mass \cite{Berenstein:2002jq}. Various properties of the heavy and light modes were studied in ~\cite{Ahn:2008aa,Zarembo:2009au,Sundin:2009zu}. Furthermore, the $AdS_{4}/CFT_{3}$ magnon dispersion relation was found to be  $\epsilon=\frac{1}{2}\sqrt{1+8h(\lambda)\sin^{2}\frac{p}{2}}$
where $h(\lambda)=\lambda$ for $\lambda \gg 1$ and $h(\lambda)=2\lambda^{2}$ for $\lambda \ll 1$. This is in contrast to the magnon dispersion relation for $AdS_{5}/CFT_{4}$, where $h(\lambda)=\frac{\sqrt{\lambda}}{4\pi}$
for all values of $\lambda$. One possible reason why the  $AdS_{4}/CFT_{3}$ magnon dispersion
receives corrections at strong coupling is that the theory only has $3/4$ maximal supersymmetry. Another consequence of the less-than-maximal supersymmetry is that the radius of $AdS_4 \times \CP$ varies as a function of $\lambda$, although this only becomes relevant at two loops in the sigma model ~\cite{Bergman:2009zh}.

Perhaps the most puzzling new feature of the $AdS_{4}/CFT_{3}$ correspondence
arises when computing the one-loop correction to the energy of classical
solutions to type IIA string theory in $AdS_{4}\times \CP$. Note that the one-loop corrections we are describing
correspond to quantum corrections to the world-sheet theory and
$\alpha'$ corrections to the classical string theory. In particular, several groups found a disagreement with the all-loop Bethe ansatz after computing the one-loop correction to the energy of the folded spinning string in $AdS_{4}\times \CP$. In computing the one-loop correction, these groups used the same prescription for adding up fluctuation frequencies that was used in $AdS_{5}\times S^{5}$ \cite{Krishnan:2008zs, McLoughlin:2008ms, Alday:2008ut}.
The authors of \cite{Gromov:2008fy} subsequently proposed an alternative summation prescription which achieves
agreement with the all-loop Bethe ansatz by treating the frequencies of heavy and light modes on unequal footing. This prescription is not applicable to type IIB string theory on $AdS_{5}\times S^{5}$ because there is no distinction between heavy and light frequencies in this theory. Hence, the prescription proposed in \cite{Gromov:2008fy} is special to the $AdS_{4}/CFT_{3}$ correspondence. Reference \cite{McLoughlin:2008he} pointed out that the discrepancy can also be resolved if one takes $\sqrt{h(\lambda)}=\sqrt{\lambda}+a_1+\mathcal{O}\left(1/\sqrt{\lambda}\right)$ with $a_1\neq 0$ when doing world-sheet calculations. Although the algebraic curve calculation in \cite{Shenderovich:2008bs} found that this correction should be zero, the authors in \cite{McLoughlin:2008he} argue that different values of $a_1$ can be consistent because $a_1$ may be scheme-dependent.

In this paper we extend the study of one-loop corrections in $AdS_{4}\times \CP$ by computing one-loop corrections for solutions with nontrivial support in $\CP$ and trivial support in $AdS_{4}$, notably a rotating point-particle and a circular string with two equal angular momenta in $\CP$, which we refer to as the spinning string. The latter solution is the $AdS_{4}/CFT_{3}$ analogue of the $SU(2)$ circular
string which was discovered in~\cite{Frolov:2003qc} and studied extensively in the $AdS_{5}/CFT_{4}$ correspondence \cite{Frolov:2003tu,Frolov:2004bh,Beisert:2005mq}. The point-particle and spinning string
solutions are especially interesting to study in the $AdS_{4}/CFT_{3}$
context because they avoid the $\kappa$-symmetry issues described
above (since they have trivial support in $AdS_{4}$). Various string solutions with support in $CP^3$ were also constructed in~\cite{Chen:2008qq,Ahn:2008hj}, however one-loop corrections were not considered in those papers.

In order to compute the one-loop correction to the energy of a classical solution, we must first compute the spectrum of fluctuations about the solution. This can be computed by expanding the Green-Schwarz (GS) action to quadratic
order in the fluctuations and finding the normal modes of the resulting equations of motion. We refer to this method
as the world-sheet (WS) approach.
Alternatively, the spectrum can be computed
from the algebraic curve corresponding to this solution using semi-classical
techniques. We refer to this as the algebraic curve (AC) approach. This approach was developed for type IIB string theory in $AdS_{5} \times S^{5}$ in~\cite{Gromov:2007aq} and then adapted to type IIA string theory in $AdS_{4} \times \CP$ in~\cite{Gromov:2008bz}.
In this paper, we compute the spectrum of fluctuations about the point-particle and spinning string using both approaches and find that the algebraic curve frequencies agree with the world-sheet frequencies up to constant shifts and shifts in mode number.

Although the algebraic curve and world-sheet spectra look very similar, they have very different properties. In particular, the algebraic curve spectrum gives a divergent one-loop correction if we use the same prescription for adding up the frequencies that was used in $AdS_{5}\times S^5$. Since the point-particle is a BPS solution we expect that its one-loop correction should vanish. Furthermore, since the spinning string solution becomes near-BPS in a certain limit, we expect its one-loop correction to be nonzero but finite. Hence the algebraic curve does not give one-loop corrections which are compatible with supersymmetry if one uses the standard summation prescription.

We propose a new summation prescription that gives a vanishing one-loop correction for the point-particle and a finite one-loop correction for the spinning string when used with both the algebraic curve spectrum and the world-sheet spectrum. This prescription has certain similarities to the one that was proposed by Gromov and Mikhaylov in~\cite{Gromov:2008fy}, however our motivation for introducing it is somewhat different. Whereas they proposed a new summation prescription in order to get a one-loop correction to the energy of the folded spinning string which agrees with the all-loop Bethe ansatz, we find that a new summation prescription is required for a much more basic reason: consistency of the algebraic curve with supersymmetry. In principle, we obtain three predictions for the one-loop correction to the spinning-string energy; one coming from the algebraic curve and two coming from the world-sheet (since the world-sheet spectrum gives finite results using both the old and new summation prescriptions). However, if we expand in the large-$\mathcal{J}$ limit (where $\mathcal{J}=\frac{J}{\sqrt{2 \pi^2 \lambda}}$ and $J$ is the spin) and evaluate the sums at each order of $\mathcal{J}$ using $\zeta$-function regularization, we find that all three predictions are the same (up to so-called non-analytic and exponentially suppressed terms which are sub-dominant). In this way we get a single prediction for the one-loop correction to the spinning string energy. Furthermore, we show that this result is consistent with the predictions of the Bethe ansatz.

The structure of this paper is as follows. In section 2, we review the world-sheet approach, the algebraic curve approach, and summation prescriptions. It should be noted that our versions of the world-sheet and algebraic curve formalisms have some new features. In particular, we recast the quadratic GS action in the $AdS_4 \times CP^{3}$ supergravity background in a way that removes half of the fermionic degrees of freedom explicitly and we reformulate the algebraic curve approach using off-shell techniques which make calculations much more efficient. In section 3.1, we present the classical solution for a point-particle rotating in $\CP$ and describe the gauge theory operator dual to this solution. In the rest of section 3 we summarize the fluctuation frequencies for the point-particle solution and compute the one-loop correction using the standard summation prescription used in $AdS_5 \times S^5$ as well as our new summation prescription.\footnote{ Although the spectrum of the point-particle was already computed using the algebraic curve in~\cite{Gromov:2008bz}, we present it again using more efficient techniques.} In section 4.1, we present
the classical solution for a spinning string with two equal angular momenta in $\CP$ and propose the gauge theory operator dual to this solution.  In the rest of section 4 we summarize the fluctuation frequencies for the spinning-string solution, analyze various properties of the one-loop correction to its energy, and make a prediction for the anomalous dimension of its dual gauge theory operator.\footnote{The authors in  \cite{Chen:2008qq} made similar conjectures for the gauge theory operators dual to the point-particle and spinning string, however the classical solutions considered in that paper have different charges than the ones constructed in this paper.} Section 5 presents our conclusions. Appendix A reviews some basic properties of the dual gauge theory. Appendices B and C review the geometry of $AdS_{4} \times \CP$ as well as our Dirac matrix conventions, and the remaining appendices provide detailed derivations of the fluctuation frequencies for the point-particle and spinning string using both the algebraic curve approach and the world-sheet approach. In Appendix H, we use the Bethe ansatz to compute the leading two contributions to the anomalous dimension of operator dual to the spinning and verify that they agree with the prediction we obtain using string theory.

\section{Review of Formalism}

\subsection{World-Sheet Formalism}

The world-sheet approach for computing the spectrum of fluctuations about a
classical solution in $AdS_{5}\times S^{5}$ was developed in~\cite%
{Frolov:2002av}. In this section we review how to compute the spectrum of
fluctuations around a classical solution to type IIA string theory in a supergravity background which consists of the following string frame
metric, dilaton, and Ramond-Ramond forms \cite{Aharony:2008ug}:
\begin{subequations}
\label{IIAGeom}
\begin{eqnarray}
ds^{2} &=&G_{MN}dx^{M}dx^{N}=R^{2}\left( \frac{1}{4}%
ds_{AdS_{4}}^{2}+ds_{CP^{3}}^{2}\right) ,  \label{IIAGeomM} \\
e^{\phi } &=&\frac{R}{k},  \label{Dilaton} \\
\,F_{4} &=&\frac{3}{8}kR^{2}{\normalcolor Vol}_{AdS_{4}},  \label{F4} \\
F_{2} &=&k\mathbf{{J},}  \label{F2}
\end{eqnarray}%
where $R^{2}$ is the radius of
curvature in string units, $\mathbf{J}$ is the Kähler form on $CP^{3}$, and $%
k$ is an integer corresponding to the level of the dual Chern-Simons theory.
Note that the $AdS_{4}$ space has radius $R/2$ while the $CP^{3}$ space has
radius $R$. The metric for a unit $AdS_{4}$ space given by
\end{subequations}
\begin{equation}
ds_{AdS_{4}}^{2}=-\cosh ^{2}\rho \mathrm{d}t^{2}+\mathrm{d}\rho ^{2}+\sinh
^{2}\rho \left( \mathrm{d}\theta ^{2}+\sin ^{2}\theta \mathrm{d}\phi
^{2}\right) ,  \label{Ads4metric}
\end{equation}%
and the metric for a unit $CP^{3}$ space is given by
\begin{eqnarray}
ds_{CP^{3}}^{2} &=&\mathrm{d}\xi ^{2}+\cos ^{2}\xi \sin ^{2}\xi \left(
\mathrm{d}\psi +\frac{1}{2}\cos \theta _{1}\mathrm{d}\varphi _{1}-\frac{1}{2}%
\cos \theta _{2}\mathrm{d}\varphi _{2}\right) ^{2}  \label{cp3} \\
&&+\frac{1}{4}\cos ^{2}\xi \left( \mathrm{d}\theta _{1}^{2}+\sin ^{2}\theta
_{1}\mathrm{d}\varphi _{1}^{2}\right) +\frac{1}{4}\sin ^{2}\xi \left(
\mathrm{d}\theta _{2}^{2}+\sin ^{2}\theta _{2}\mathrm{d}\varphi
_{2}^{2}\right) ,  \notag
\end{eqnarray}%
where $0\leq \xi <\pi /2,$ $0 \leq \psi < 2\pi ,$ $0\leq \theta _{i}\leq \pi ,$ and $0\leq \varphi _{i}<2\pi .$ More details about the geometry of $%
AdS_{4}\times CP^{3}$ are given in Appendix B.

Using the metric in Eq.~(\ref{IIAGeom}), the bosonic part of the string
Lagrangian in conformal gauge is given by
\begin{equation}
\ \mathcal{L}_{bose}=\frac{1}{4\pi }\eta ^{ab}G_{MN}\partial
_{a}X^{M}\partial _{b}X^{N},  \label{eq:bosaction}
\end{equation}%
where $a,b=\tau ,\sigma $ are world-sheet indices, $\eta ^{ab}={\normalcolor %
diag}\left[ -1,1\right] $, and we have set $\alpha ^{\prime }=1$. Because $%
AdS_{4}$ has two Killing vectors and $CP^{3}$ has three Killing vectors, any
solution to the bosonic equations of motion has at least five conserved charges. In
particular, the two $AdS_{4}$ charges are given by
\begin{equation}
E=\sqrt{{\lambda /2}}\int_{0}^{2\pi }d\sigma \cosh ^{2}\rho \dot{t}{,}
\label{eq:E}
\end{equation}%
\begin{equation}
S=\sqrt{{\lambda /2}}\int_{0}^{2\pi }d\sigma \sinh ^{2}\rho \sin ^{2}\theta
\dot{\phi}.  \label{SpinS}
\end{equation}%
and the three $CP^{3}$ charges are given by
\begin{subequations}
\label{Js}
\begin{eqnarray}
J_{\psi } &=&2\sqrt{{2\lambda}}\int_{0}^{2\pi }d\sigma \cos ^{2}\xi \sin
^{2}\xi \left( \dot{\psi}+\frac{1}{2}\cos \theta _{1}\dot{\phi}_{1}-\frac{1}{%
2}\cos \theta _{2}\dot{\phi}_{2}\right) ,  \label{eq:Jpsi1} \\
J_{\phi _{1}} &=&\sqrt{{\lambda /2}}\int_{0}^{2\pi }d\sigma \cos ^{2}\xi
\sin ^{2}\theta _{1}\dot{\phi}_{1}  \label{eq:JPhi1} \\
&&+\sqrt{{2\lambda }}\int_{0}^{2\pi }d\sigma \cos ^{2}\xi \sin ^{2}\xi
\left( \dot{\psi}+\frac{1}{2}\cos \theta _{1}\dot{\phi}_{1}-\frac{1}{2}\cos
\theta _{2}\dot{\phi}_{2}\right) \cos \theta _{1},  \notag \\
J_{\phi _{2}} &=&\sqrt{{\lambda /2}}\int_{0}^{2\pi }d\sigma \sin ^{2}\xi
\sin ^{2}\theta _{2}\dot{\phi}_{2}  \label{eq:JPhi2} \\
&&+\sqrt{{2\lambda }}\int_{0}^{2\pi }d\sigma \cos ^{2}\xi \sin ^{2}\xi
\left( \dot{\psi}+\frac{1}{2}\cos \theta _{1}\dot{\phi}_{1}-\frac{1}{2}\cos
\theta _{2}\dot{\phi}_{2}\right) \cos \theta _{2,}  \notag
\end{eqnarray}%
where $E$ is the energy and $S$, $J_{\psi }$, $J_{\phi _{1}}$, and $J_{\phi
_{2}}$ are angular momenta.

A solution to the bosonic equations of motion is said to be a classical
solution if it also satisfies the Virasoro constraints
\end{subequations}
\begin{equation}
\ G_{MN}\left( \partial _{\tau }X^{M}\partial _{\tau }X^{N}+\partial
_{\sigma }X^{M}\partial _{\sigma }X^{N}\right) =0,\,\,\,G_{MN}\partial
_{\tau }X^{M}\partial _{\sigma }X^{N}=0{\normalcolor.}  \label{eq:Virasoro}
\end{equation}%
Note that these are the only constraints that relate motion in $AdS_{4}$ to
motion in $CP^{3}$.

The spectrum of bosonic fluctuations around a classical solution can be
computed by expanding the bosonic Lagrangian in Eq.~(\ref{eq:bosaction}) to
quadratic order in the fluctuations and finding the normal modes of the
resulting equations of motion. In the examples we consider, we find that two of the
bosonic modes are massless and the other eight are massive. While the eight
massive modes correspond to the physical transverse degrees of freedom, the
two massless modes can be discarded. One way to see that the massless modes
can be discarded is by expanding the Virasoro constraints to linear order in
the fluctuations \cite{Frolov:2002av}.

To compute the spectrum of fermionic fluctuations, we only need the
quadratic part of the fermionic GS action for type IIA string theory. This
action describes two 10-dimensional
Majorana-Weyl spinors of opposite chirality which can be combined into a
single non-chiral Majorana spinor $\Theta $. The quadratic GS action for type IIA string theory in a general background
can be found in \cite{Cvetic:1999zs}. For the supergravity background in
Eq.~(\ref{IIAGeom}), the quadratic Lagrangian for the fermions is given by
\begin{equation}
\mathcal{L}_{fermi}=\bar{\Theta}\left( \eta ^{ab}-\epsilon ^{ab}\Gamma _{11}\right)
e_{a}\left[ (\partial _{b}+\frac{1}{4}\omega _{b})+\frac{1}{8}e^{\phi
}\left( -\Gamma _{11}\Gamma \cdot F_{2}+\Gamma \cdot F_{4}\right) e_{b}%
\right] \Theta {,}  \label{eq:GS}
\end{equation}%
where $\bar{\Theta}=\Theta ^{\dagger }\Gamma ^{0}$, $\epsilon ^{\tau \sigma
}=-\epsilon ^{\sigma \tau }=1$, $e_{a}=\partial _{a}X^{M}e_{M}^{A}\Gamma _{A}
$, $\omega _{a}=\partial _{a}X^{M}\omega _{M}^{AB}\Gamma _{AB}$, and $\Gamma
\cdot F_{(n)}=\frac{1}{n!}\Gamma ^{N_{1}...N_{n}}F_{N_{1}...N_{n}}$. Note
that $M$ is a base-space index while $A,B=0,...,9$ are tangent-space
indices. Explicit formulas for $e_{M}^{A}$, $\omega _{M}^{AB}$, $\Gamma
\cdot F_{2}$, and $\Gamma \cdot F_{4}$ are provided in Appendix B. Explicit
formulas for the Dirac matrices are provided in Appendix C.

We will now recast the fermionic Lagrangian in Eq.~(\ref{eq:GS}) in form that allows us to compute the fermionic fluctuation frequencies in a straightforward way. First we note that after rearranging terms, Eq.~(\ref{eq:GS}) can be written as
\begin{equation}
\frac{\mathcal{L}_{fermi}}{2K}=-\bar{\Theta}_{+}\Gamma _{0}\left[ \partial
_{\tau}-\Gamma _{11}\partial _{\sigma}+\frac{1}{4}\left( \omega _{\tau}-\Gamma
_{11}\omega _{\sigma}\right)\right]\Theta-2K\bar{\Theta}_{+}\Gamma _{0}\Gamma \cdot F\Gamma _{0}\Theta_{+} ,  \label{GS2}
\end{equation}%
where we define $K = \partial _{\tau }X^{M}e_{M}^{0}$, $\Theta
_{+}=P_{+}\Theta $, and%
\begin{equation}
P_{+}= -\frac{1}{2K}\Gamma _{0}\left( e_{\tau}+e_{\sigma}\Gamma _{11}\right)
,  \label{Proj}
\end{equation}
\begin{equation}
\Gamma \cdot F =\frac{1}{8}e^{\phi }\left( -\Gamma _{11}\Gamma \cdot
F_{2}+\Gamma \cdot F_{4}\right) \label{GammaF}.
\end{equation}%
Note that $P_{+}=P_{+}^{\dag }$ and if
the classical solution satisfies
\begin{equation}
\partial _{\sigma }X^{M }e_{M}^{0}=0,  \label{ct1}
\end{equation}
then $P_{+}$ is a projection operator, i.e. $P_{+}^{2}=P_{+}$. In addition, if the classical solution satisfies
\begin{equation}
P_{+}\left[ P_{+},\omega _{\tau }-\Gamma _{11}\omega _{\sigma }\right] =0,
\label{ct2}
\end{equation}%
then the fermionic Lagrangian simplifies to
\begin{equation}
\frac{\mathcal{L}_{fermi}}{2K}=-\bar{\Theta}_{+}\Gamma _{0}\left[ \partial
_{\tau}-\Gamma _{11}\partial _{\sigma}+\frac{1}{4}\left( \omega _{\tau}-\Gamma
_{11}\omega _{\sigma}\right) +2K\left( \Gamma \cdot F\Gamma _{0}\right) %
\right] \Theta _{+}.  \label{GS22}
\end{equation}
Finally, if we consider the Fourier mode $\Theta \left( \sigma ,\tau \right) =%
\tilde{\Theta} \exp \left( -i\omega \tau +in\sigma \right) $, where $%
\tilde{\Theta}$ is a constant spinor, then the equations of motion for the
fermionic fluctuations are given by%
\begin{equation}
\left\{ P_{+}\left[ i\omega +in\Gamma _{11}-\frac{1}{4}\left( \omega
_{\tau}-\Gamma _{11}\omega _{\sigma}\right) -2K\left( \Gamma \cdot F\Gamma
_{0}\right) \right] P_{+}\right\} \tilde{\Theta}=0.  \label{Feom}
\end{equation}
One can choose a basis where $P_{+}$ has the form $\left(\begin{array}{cc}
1 & 0\\
0 & 0\end{array}\right)$ (where each element in the $2\times2$ matrix corresponds to a $16\times16$
matrix). In this basis, the matrix on the left-hand side of Eq.~(\ref{Feom}) will
have the form $\left(\begin{array}{cc}
{\bf A} & 0\\
0 & 0\end{array}\right)$. The fermionic frequencies are determined by taking the determinant
of ${\bf A}$ and finding its roots.

Only half of the fermionic components appear in the
Lagrangian in Eq.~(\ref{GS22}). Hence, a natural choice for fixing kappa-symmetry is to set the
other components to zero by imposing the gauge condition $\Theta =\Theta _{+}
$. This gives the desired number of fermionic degrees of freedom. In
particular, before imposing the Majorana condition, $\Theta $ has 32 complex
degrees of freedom. When the classical solution satisfies Eqs. (\ref{ct1}) and (\ref{ct2}), the quadratic GS action can be recast in terms of
projection operators that remove half of $\Theta $'s components, leaving
16 complex degrees of freedom. After solving the fermionic equations of
motion, one then finds that only half the solutions have positive energy, leaving
eight complex degrees of freedom. Finally, after imposing the Majorana
condition we should be left with eight real degrees of freedom, which
matches the number of transverse bosonic degrees of freedom.
Explicit calculations of the fermionic frequencies for the classical
solutions studied in this paper are described in Appendices D.2 and F.2.


\subsection{Algebraic Curve Formalism}

The procedure for computing the spectrum of excitations about a classical
string solution using the $AdS_{4}/CFT_{3}$ algebraic curve was first
presented in~\cite{Gromov:2008bz}. In this section, we reformulate this
procedure in terms of an off-shell formalism similar to the one that was
developed for the $AdS_{5}/CFT_{4}$ algebraic curve in~\cite{Gromov:2008ec}.
The off-shell formalism makes things much more efficient. First we describe
how to construct the classical algebraic curve. Then we describe how to
semi-classically quantize the curve and obtain the spectrum of excitations.

\subsubsection{Classical Algebraic Curve}

For type IIA string theory in $AdS_{4}\times CP^{3}$, any classical solution
can be encoded in a 10-sheeted Riemann surface whose branches, called
quasi-momenta, are denoted by
\begin{equation*}
\left\{q_{1},q_{2},q_{3},q_{4},q_{5},q_{6},q_{7},q_{8},q_{9},q_{10}\right\}.
\end{equation*}
This algebraic curve corresponds to the fundamental representation of $OSp(6|4)$, which is ten-dimensional.
Furthermore, the quasimomenta are not all independent. In particular
\begin{equation}
\left( q_{1}(x),q_{2}(x),q_{3}(x),q_{4}(x),q_{5}(x)\right) =-\left(
q_{10}(x),q_{9}(x),q_{8}(x),q_{7}(x),q_{6}(x)\right) ,  \label{eq:3}
\end{equation}%
where $x$ is a complex number called the spectral parameter. To compute the quasimomenta, it is
useful to parameterize $AdS_{4}$ and $\CP$ using the following embedding coordinates
\begin{eqnarray*}
n_{1}^{2}+n_{2}^{2}-n_{3}^{2}-n_{4}^{2}-n_{5}^{2} &=&1 \\
\sum_{I=1}^{4}\left\vert z^{I}\right\vert ^{2} &=&1,\qquad z^{I}\sim
e^{i\lambda }z^{I},
\end{eqnarray*}%
where $\lambda \in {\mathbb{R}}$.
A classical solution in the global coordinates of Eqs.~(\ref{Ads4metric}) and (\ref{cp3}) can be converted to embedding coordinates using Eqs.~(\ref{eq:adsmap}) and (\ref{eq:mapping}) provided in Appendix B. One can then compute the following connection:
\begin{equation}
j_{a}(\tau ,\sigma )=2\left(
\begin{array}{cc}
n_{i}\partial _{a}n_{j}-n_{j}\partial _{a}n_{i} & 0 \\
0 & z_{I}^{\dagger }D _{a}z^{J}-z^{J}D_{a}z_{I}^{\dagger }%
\end{array}%
\right) {,}  \label{eq:curr}
\end{equation}%
%
where $a \in \left\{\tau ,\sigma \right\} $, $D_{a}=\partial_{a}+i A_{a}$, and $A_{a}=i \sum_{I=1}^{4} z_I^{\dagger} \partial_{a}z^{I}$~\cite{Gromov:2008bz}. This connection is a $9\times 9$ matrix and
transforms under the bosonic part of the supergroup $OSp(6|4)$, notably $SU(4)\times SO(3,2)\sim O(6)\times Sp(4)$. A key property is that it is flat, which allows us to construct the following monodromy matrix:
\begin{equation}
\Lambda (x)=P\exp \frac{1}{x^{2}-1}\int_{0}^{2\pi }d\sigma \left[ j_{\sigma
}(\tau ,\sigma )+xj_{\tau }(\tau ,\sigma )\right],  \label{eq:mono}
\end{equation}%
where $P$ is the path-ordering symbol and the integral is over a loop of
constant world-sheet time $\tau $. It can be shown that the eigenvalues of $%
\Lambda (x)$ are independent of $\tau $.

The quasimomenta are related to the eigenvalues of the monodromy matrix. In
particular, if we diagonalize the monodromy matrix we will find that the eigenvalues of the $AdS_{4}$ part are in general given by%
\begin{equation}
\left\{ e^{i\hat{p}_{1}(x)},e^{i\hat{p}_{2}(x)},e^{i\hat{p}_{3}(x)},e^{i\hat{%
p}_{4}(x)},1\right\} ,  \label{eq:hat}
\end{equation}%
where $\hat{p}_{1}(x)+\hat{p}_{4}(x)=\hat{p}_{2}(x)+\hat{p}_{3}(x)=0$, while
the eigenvalues from the $CP^{3}$ part are given by%
\begin{equation}
\left\{ e^{i\tilde{p}_{1}(x)},e^{i\tilde{p}_{2}(x)},e^{i\tilde{p}%
_{3}(x)},e^{i\tilde{p}_{4}(x)}\right\} ,  \label{eq:tilde}
\end{equation}%
where $\sum_{i=1}^{4}\tilde{p}_{i}(x)=0$. The classical quasimomenta are
then defined as
\begin{equation}
\left( q_{1},q_{2},q_{3},q_{4},q_{5}\right) =\left( \frac{\hat{p}_{1}+\hat{p}%
_{2}}{2},\frac{\hat{p}_{1}-\hat{p}_{2}}{2},\tilde{p}_{1}+\tilde{p}_{2},%
\tilde{p}_{1}+\tilde{p}_{3},\tilde{p}_{1}+\tilde{p}_{4}\right) ,
\label{eq:pq}
\end{equation}%
where we have suppressed the $x$-dependence. From this formula, we see that $%
q_{1}(x)$ and $q_{2}(x)$ correspond to the $AdS_{4}$ part of the algebraic
curve, while $q_{3}(x)$, $q_{4}(x)$, and $q_{5}(x)$ correspond to the $CP^{3}$
part of the algebraic curve.

\subsubsection{Semi-Classical Quantization}

\begin{figure}[tb]
\center
\includegraphics{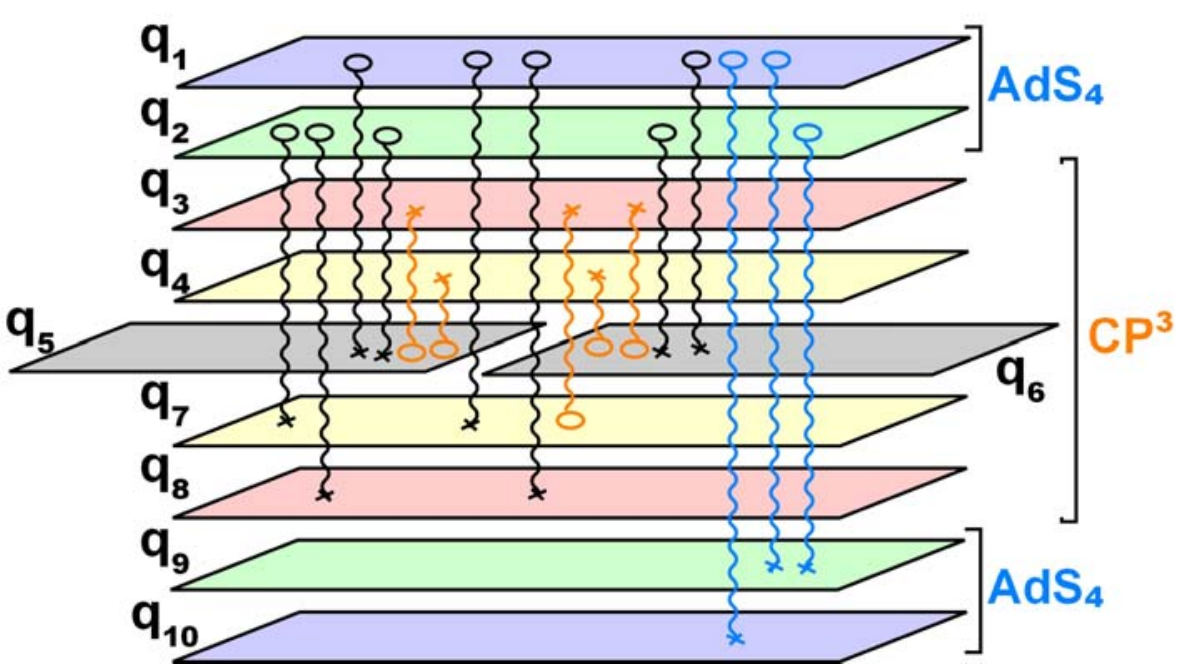}
\caption{Depiction of the fluctuations of the $AdS_{4}\times CP^{3}$ algebraic
curve. Each fluctuation corresponds to a pole which connects two sheets.}
\label{F1}
\end{figure}

The algebraic curve will generically have cuts connecting several pairs of
sheets. These cuts encode the classical physics. To perform semiclassical
quantization, we add poles to the algebraic curve which correspond to
quantum fluctuations. Each pole connects two sheets. In particular the bosonic
fluctuations connect two $AdS$ sheets or two $CP^3$ sheets and the fermionic
fluctuations connect an $AdS$ sheet to a $CP^3$ sheet. See Fig.~(\ref{F1}) for a depiction of the fluctuations. In total there are eight
bosonic and eight fermionic fluctuations and they are labeled by the pairs
of sheets that their poles connect. The labels are referred to as
polarizations and are summarized in Table~\ref{T1}.

\begin{table}[th]
\caption{Labels for the fluctuations (\textcolor{red}{\bf heavy},\textcolor{blue}{\bf light}) of the $AdS_{4}\times \CP$ algebraic curve.  }
\label
{T1}
\begin{equation}
\begin{array}{c|c}
\toprule & \mathrm{\mathbf{Polarizations~(i,j)}} \\
\midrule {\rm \bf AdS} &
\begin{array}{l}
\textcolor{red}{\bf{(1,10/1,10);(2,9/2,9);(1,9/2,10)}}%
\end{array}
\\
\midrule {\rm \bf Fermions} &
\begin{array}{l}
\textcolor{red}{\bf{(1,7/4,10);(1,8/3,10);(2,7/4,9);(2,8/3,9)}} \\
\textcolor{blue}{\bf{(1,5/6,10);(1,6/5,10);(2,5/6,9);(2,6/5,9)}}%
\end{array}
\\
\midrule CP^{3} &
\begin{array}{l}
\textcolor{red}{\bf{(3,7/4,8)}} \\
\textcolor{blue}{\bf{(3,5/6,8);(3,6/5,8);(4,5/6,7);(4,6/5,7)}}%
\end{array}
\\
\bottomrule
\end{array}
\notag
\end{equation}%
\end{table}
Notice that every fluctuation can be labeled by two equivalent
polarizations because every pole
connects two equivalent pairs of sheets as a consequence of Eq.~(\ref{eq:3}). Fluctuations connecting sheet 5 or 6 to any other sheet are defined to be light. Notice
that there are eight light excitations. All the others are defined to be
heavy excitations. The physical significance of this terminology will become
clear later on. When we compute the spectrum of fluctuations about the point
particle in section 3 for example, we will find that the heavy excitations
are twice as massive as the light excitations.

When adding poles, we must take into account the level-matching condition
\begin{equation}
\sum_{n=-\infty }^{\infty }n\sum_{ij}N_{n}^{ij}=0,  \label{eq:11}
\end{equation}%
where $N_{n}^{ij}$ is the number of excitations with polarization $ij$ and
mode number $n$. Furthermore, the locations of the poles are not arbitrary;
they are determined by the following equation:
\begin{equation}
q_{i}\left( x_{n}^{ij}\right) -q_{j}\left( x_{n}^{ij}\right) =2\pi n,
\label{eq:8}
\end{equation}%
where $x_{n}^{ij}$ is the location of a pole corresponding to a fluctuation
with polarization $ij$ and mode number $n$. 

In addition to adding poles to the algebraic curve, we must also add fluctuations to the classical
quasimomenta. These fluctuations will depend on the spectral parameter $x$
as well as the locations of the poles, which we will denote by the
collective coordinate $y$. The functional form of the fluctuations is
determined by some general constraints:
\begin{itemize}
\item They are not all independent:%
\begin{equation*}
\left(%
\begin{array}{c}
\delta q_{1}(x,y) \\
\delta q_{2}(x,y) \\
\delta q_{3}(x,y) \\
\delta q_{4}(x,y) \\
\delta q_{5}(x,y)%
\end{array}%
\right)=-\left(%
\begin{array}{c}
\delta q_{10}(x,y) \\
\delta q_{9}(x,y) \\
\delta q_{8}(x,y) \\
\delta q_{7}(x,y) \\
\delta q_{6}(x,y)%
\end{array}%
\right){\normalcolor .}
\end{equation*}

\item They have poles near the points $x=\pm1$ and the residues of these
poles are synchronized as follows:
\begin{equation}
\lim_{x\rightarrow\pm1}\left(\delta q_{1}(x,y),\delta q_{2}(x,y),\delta
q_{3}(x,y),\delta q_{4}(x,y),\delta q_{5}(x,y)\right)\propto\frac{1}{x\pm1}%
\left(1,1,1,1,0\right){\normalcolor .}  \label{eq:synch}
\end{equation}

\item There is an inversion symmetry:
\begin{equation}
\left(%
\begin{array}{c}
\delta q_{1}(1/x,y) \\
\delta q_{2}(1/x,y) \\
\delta q_{3}(1/x,y) \\
\delta q_{4}(1/x,y) \\
\delta q_{5}(1/x,y)%
\end{array}%
\right)=\left(%
\begin{array}{c}
-\delta q_{2}(x,y) \\
-\delta q_{1}(x,y) \\
-\delta q_{4}(x,y) \\
-\delta q_{3}(x,y) \\
\delta q_{5}(x,y)%
\end{array}%
\right){\normalcolor .}  \label{inversion}
\end{equation}

\item The fluctuations have the following large-$x$ behavior:
\begin{equation}
\lim_{x\rightarrow \infty }\left(
\begin{array}{c}
\delta q_{1}(x,y) \\
\delta q_{2}(x,y) \\
\delta q_{3}(x,y) \\
\delta q_{4}(x,y) \\
\delta q_{5}(x,y)%
\end{array}%
\right) \simeq \frac{1}{2gx}\left(
\begin{array}{c}
\Delta (y)+N_{19}+2N_{1\,10}+N_{15}+N_{16}+N_{17}+N_{18} \\
\Delta (y)+2N_{29}+N_{2\,10}+N_{25}+N_{26}+N_{27}+N_{28} \\
-N_{35}-N_{36}-N_{37}-N_{39}-N_{3\,10} \\
-N_{45}-N_{46}-N_{48}-N_{49}-N_{4\,10} \\
N_{35}+N_{45}-N_{57}-N_{58}-N_{15}-N_{25}+N_{59}+N_{5\,10}%
\end{array}%
\right) ,  \label{eq:asymptotics}
\end{equation}%
where $g=\sqrt{\lambda /8}$, $N_{ij}=\sum_{n=-\infty }^{\infty }N_{n}^{ij}$,
and $\Delta (y)$ is called the anomalous part of the energy shift. Whereas
the $N_{n}^{ij}$ are inputs of the calculation, $\Delta (y)$ will be
determined in the process of determining the fluctuations of the
quasi-momenta. The factor of two that appears in front of $N_{1\,10}$ and $N_{29}$ is a consequence of the symmetry in Eq.~(\ref{eq:3}).
The coefficients of the other terms on the right hand side of equation Eq.~(\ref{eq:asymptotics}) can be determined using arguments similar to those in \cite{Gromov:2007aq}.

\item Finally, when the spectral parameter approaches the location of one of
the poles, the fluctuations have the following form:
\begin{equation}
\lim_{x\rightarrow x_{n}^{ij}}\delta q_{k}\propto \frac{\alpha
(x_{n}^{ij})N_{n}^{ij}}{x-x_{n}^{ij}},\,\,\,\alpha (x)=\frac{1}{2g}\frac{%
x^{2}}{x^{2}-1},  \label{eq:alpha}
\end{equation}%
where the proportionality constants can be read off from the coefficient of
$N_{ij}$ in the $k$'th row of Eq.~(\ref{eq:asymptotics}).
\end{itemize}

After computing the anomalous part of the energy shift, the
fluctuation frequency is given by
\begin{equation}
\Omega (y)=\Delta (y)+\sum_{AdS_{4}}N^{ij}+\frac{1}{2}\sum_{ferm}N^{ij}{%
\normalcolor.}  \label{eq:9}
\end{equation}%
It is useful to consider the fluctuation frequency without fixing the value
of $y$. In this case, the fluctuation frequency is said to be off-shell.

Using arguments similar to those in \cite{Gromov:2008ec}, we find all the relations among the off-shell frequencies.
First, all the light off-shell frequencies are related by
\begin{equation}
\label{LL}
\Omega_{i6}(y)=\Omega _{i5}(y)
\end{equation}
where $i=1,2,3,4$.

Second, all the heavy off-shell frequencies can be written as the sum of two light off-shell frequencies as summarized in table \ref{tLH}.
\begin{table}[th]
\caption{Relations between heavy and light off-shell frequencies. }
\label{tLH}
\begin{equation*}
\begin{array}{cc|c}
\toprule \textcolor{red}{\bf{Heavy}} & \textcolor{blue}{\bf{Light}} &  \\
\midrule
\begin{array}{r}
\Omega _{29} =  \\
\Omega _{1~10}=   \\
\Omega _{19}=     \\
\end{array}
&
\begin{array}{l}
2\Omega _{25}   \\
2\Omega _{15}  \\
\Omega _{15}+\Omega _{25}
\end{array}
&
{\rm \bf AdS}
\\
\midrule
\begin{array}{l}
\Omega _{27}= \\
\Omega _{17}= \\
\Omega _{28}= \\
\Omega _{18}=
\end{array}
&
\begin{array}{l}
\Omega _{25}+\Omega _{45}  \\
\Omega _{15}+\Omega _{45} \\
\Omega _{25}+\Omega _{35}   \\
\Omega _{15}+\Omega _{35}
\end{array}
&
{\rm \bf Fermions}
\\
\midrule
\begin{array}{l}
\Omega _{37}=\\
\end{array}
&
\begin{array}{l}
\Omega _{35}+\Omega _{45}
\end{array}
&
{\mathbb C \mathbb P}^3
\\
\bottomrule
\end{array}
\end{equation*}%
\end{table}

Finally, any off-shell frequency $\Omega _{ij}$ is related to its mirror
off-shell frequency $\Omega _{\overline{ij}}$ by
\[
\Omega _{ij}\left( y\right) =-\Omega _{\overline{ij}}\left( 1/y\right)
+\Omega _{\overline{ij}}\left( 0\right) +C,
\]%
where $C=1,1/2$, or $0$ for AdS, Fermionic, or $CP^{3}$ polarizations respectively. The
mirror polarization $\left( \overline{i,j}\right) $ of the polarization $%
\left( i,j\right) $ can be readily found using Eq.~(\ref{inversion}), e.g.  $\left( \overline{1,10}\right)=\left( 2,9\right) ,$ $
\left( \overline{2,5}\right)= \left( 1,5\right) ,$ $\left( \overline{4,5}\right)= \left( 3,5\right) ,$ $\left( \overline{3,7}\right)= \left( 3,7\right) ,$ etc.
Using these relations, only two of the
eight light off-shell frequencies are independent. For example,
\begin{subequations}
\label{M35}
\begin{eqnarray}
\Omega _{35}\left( y\right)  &=&-\Omega _{45}\left( 1/y\right) +\Omega
_{45}\left( 0\right) , \\
\Omega _{25}\left( y\right)  &=&-\Omega _{15}\left( 1/y\right) +\Omega
_{15}\left( 0\right) +1/2.
\end{eqnarray}%
\end{subequations}
In conclusion, if we compute the off-shell frequencies $\Omega _{15}$ and $%
\Omega _{45}$, then we can determine all the other off-shell frequencies
automatically from the relations in Eqs.~(\ref{LL},\ref{M35}) and table \ref{tLH}.

The on-shell frequencies are then obtained by evaluating the off-shell frequencies at the location of the
poles which are determined by solving Eq.~(\ref{eq:8}), i.e. $\omega_{n}^{ij}=\Omega _{ij}\left( x_{n}^{ij}\right) $. It will be convenient to
organize them into the following linear combinations:
\begin{equation}
\omega _{L}(n)=\omega _{n}^{35}+\omega _{n}^{36}+\omega _{n}^{45}+\omega
_{n}^{46}-\omega _{n}^{15}-\omega _{n}^{16}-\omega _{n}^{25}-\omega _{n}^{26}
\label{eq:light}
\end{equation}%
\begin{equation}
\omega _{H}(n)=\omega _{n}^{19}+\omega _{n}^{29}+\omega _{n}^{1\,10}+\omega
_{n}^{37}-\omega _{n}^{17}-\omega _{n}^{18}-\omega _{n}^{27}-\omega _{n}^{28}
\label{eq:heavy}
\end{equation}%
where $L$ stands for light and $H$ stands for heavy. It should be noted that
heavy and light frequencies are not as well-defined in the world-sheet
approach. In general, the only way to identify heavy and light frequencies
in the world-sheet approach is by comparing the world-sheet spectrum to the
algebraic curve spectrum, i.e. a world-sheet frequency is said to be
heavy/light if the corresponding algebraic curve frequency is heavy/light.

\subsection{Summation Prescriptions}

Given the spectrum of fluctuations about a classical string solution, we
compute the one-loop correction to the string energy by adding up the
spectrum. The standard formula is
\begin{equation}
\delta E_{1-loop,old}=\lim_{N\rightarrow \infty }\frac{1}{2\kappa}\sum_{n=-N}^{N}%
\left( \sum_{i=1}^{8}\omega _{n,i}^{B}-\sum_{i=1}^{8}\omega
_{n,i}^{F}\right) ,  \label{eq:bad1}
\end{equation}%
where $\kappa$ is proportional to the classical energy (the exact formula is given in sections 3 and 4), $B/F$ stands for bosonic/fermionic, $n$ is the mode number, and $i$ is
some label. For example, if we are dealing with frequencies
computed from the algebraic curve, then they will be labeled by a pair of
integers called a polarization, as explained in section 2.2. Although this
formula works well for string solutions in $AdS_{5}\times S^{5}$, it gives a
one-loop correction which disagrees with the all-loop Bethe ansatz when
applied to the folded-spinning string in $AdS_{4}\times CP^{3}$. In~\cite{Gromov:2008fy} Gromov and Mikhaylov subsequently proposed the following formula for
computing one-loop corrections in $AdS_{4}\times CP^{3}$:
\begin{equation}
\delta E_{1-loop,{\rm GM}}=\lim_{N\rightarrow \infty }\frac{1}{2\kappa}%
\sum_{n=-N}^{N}K_{n},\,\,\,K_{n}=\left\{
\begin{array}{c}
\omega ^{H}(n)+\omega^{L}(n/2)\,\,\,n\in even \\
\omega^{H}(n)\,\,\,n\in odd%
\end{array}%
\right. ,  \label{eq:vic}
\end{equation}%
where $\omega _{n}^{L}$/$\omega _{n}^{H}$ are referred to as heavy/light
frequencies and are defined in Eqs.~(\ref{eq:light}) and (\ref{eq:heavy}).
For later convenience, we note that Eq.~(\ref{eq:bad1}) can be written in
terms of heavy and light frequencies as follows:
\begin{equation}
\delta E_{1-loop,old}=\lim_{N\rightarrow \infty }\frac{1}{2\kappa}\sum_{-N}^{N}\left(
\omega _{L}(n)+\omega _{H}(n)\right) {\normalcolor.}  \label{eq:badsum}
\end{equation}%
In the large-$\kappa$ limit, Eq.~(\ref{eq:vic}) can be approximated as the following integral:
\begin{equation}
\delta E_{1-loop}\approx \lim_{N\rightarrow \infty }\frac{1}{2\kappa}\int_{-N }^{N }\left( \omega
_{H}(n)+\frac{1}{2}\omega _{L}(n/2)\right) dn{\normalcolor.}
\label{eq:intvic}
\end{equation}%
In ~\cite{Gromov:2008fy} it was shown that Eq.~(\ref{eq:intvic}) gives a one-loop correction which agrees with the all-loop
Bethe ansatz when applied to the spectrum of the folded spinning string.

In this paper we propose a new summation prescription:
\begin{equation}
\delta E_{1-loop,new}=\lim_{N\rightarrow \infty }\frac{1}{2\kappa}\sum_{-N}^{N}\left(2\omega_{H}(2n)+\omega_{L}(n)\right){%
\normalcolor.}  \label{eq:goodsum}
\end{equation}%
This sum can be motivated physically using the observation in~\cite{Gromov:2008fy} that heavy modes with mode number $2n$ can be thought of as bound states of two light modes with mode number $n$. This suggests that only heavy modes with even mode number should contribute to the one-loop correction. The formula for the one-loop correction should therefore have the form
\[\delta E_{1-loop,new}=\lim_{N\rightarrow \infty }\frac{1}{2\kappa}\sum_{-N}^{N}\left(A\omega_{H}(2n)+B\omega_{L}(n)\right){\normalcolor .}\]
The coefficients $A$ and $B$ can then be fixed uniquely by requiring that the integral approximation to this formula reduces to Eq.~(\ref{eq:intvic}) in the large-$\kappa$ limit, ensuring that this summation prescription gives a one-loop correction to the folded spinning string energy which agrees with the all-loop Bethe ansatz. One then finds that $A=2$ and $B=1$.

One virtue of the new summation prescription in Eq.~(\ref{eq:goodsum}) compared to the one in Eq.~(\ref{eq:vic}) is that it gives more well-defined results for one-loop corrections. For example, consider the case where
$\omega_{L}(n)=-2\omega_{H}(n)=C$, where $C$ is some constant (the AC frequencies for the point-particle will have this form). In this case,
Eq.~(\ref{eq:vic}) does not have a well-defined $N\rightarrow\infty$ limit;
in particular the sum alternates between $\pm C/(4\kappa)$ depending on whether $N$ is
even or odd. On the other hand, Eq.~(\ref{eq:goodsum}) vanishes for all $N$.

\section{Point-Particle}

\subsection{Classical Solution and Dual Operator}

In terms of the coordinates of Eqs.~(\ref{Ads4metric}) and (\ref{cp3}), the
solution for a point-particle rotating with angular momentum $J$ in $CP^{3}$ is given
by
\begin{equation}
t=\kappa \tau ,\,\,\,\rho=0,\,\,\,\xi =\pi
/4,\,\,\,\theta _{1}=\theta _{2}=\pi /2,\,\,\,\psi =\mathcal{J}\tau
,\,\,\,\phi _{1}=\phi _{2}=0,  \label{eq:partglob}
\end{equation}%
where $\mathcal{J}=\frac{J}{4\pi g}$ and $g=\sqrt{\lambda /8}$. This version
of the solution will be useful for doing calculations in the world-sheet
formalism. Alternatively, we can write this solution in embedding
coordinates by plugging Eq.~(\ref{eq:partglob}) into Eqs.~(\ref{eq:adsmap})
and (\ref{eq:mapping}):
\begin{equation}
n_{1}=\cos \kappa \tau ,\,\,\,n_{2}=\sin \kappa \tau
,\,\,\,n_{3}=n_{4}=n_{5}=0,\,\,\,z^{1}=z^{2}=z_{3}^{\dagger }=z_{4}^{\dagger
}=\frac{1}{2}e^{i\mathcal{J}\tau /2}{\normalcolor.}  \label{eq:partemb}
\end{equation}%
This version of the solution will be useful for doing calculations in the
algebraic curve formalism. The energy and angular momenta of the particle can be read
off from Eqs.~(\ref{eq:E}-\ref{Js}): $E=4\pi g\kappa $, $S=0$, $J_{\psi }=J$, $%
J_{\phi _{1}}=J_{\phi _{2}}=0$. Furthermore, the Virasoro constraints in
Eq.~(\ref{eq:Virasoro}) give $\kappa =\mathcal{J}$, or equivalently $E=J$.
Note that this is a BPS condition. We therefore expect that the dimension of
the dual gauge theory operator should be protected by supersymmetry.

The gauge theory operator dual to the point-particle
rotating in $CP^{3}$ should have the form
\begin{equation*}
\mathcal{O}={\rm tr}\left[\left(Z^{1}Z_{3}^{\dagger}\right)^{J}\right]{\normalcolor .}
\end{equation*}
This can be understood heuristically by associating the scalars $%
Z^{1}$,$Z^{2}$, $Z^{3}$, $Z^{4}$ with the embedding coordinates $z^{1}$, $%
z^{2}$, $z^{3}$, $z^{4}$ and noting that%
\begin{equation*}
\frac{1}{2}\left(%
\begin{array}{c}
e^{i\mathcal{J}\tau/2} \\
e^{i\mathcal{J}\tau/2} \\
e^{-i\mathcal{J}\tau/2} \\
e^{-i\mathcal{J}\tau/2}%
\end{array}%
\right)=\left(%
\begin{array}{cccc}
1/\sqrt{2} & -1/\sqrt{2} & 0 & 0 \\
1/\sqrt{2} & 1/\sqrt{2} & 0 & 0 \\
0 & 0 & 1/\sqrt{2} & -1/\sqrt{2} \\
0 & 0 & 1/\sqrt{2} & 1/\sqrt{2}%
\end{array}%
\right)\frac{1}{\sqrt{2}}\left(%
\begin{array}{c}
e^{i\mathcal{J}\tau/2} \\
0 \\
e^{-i\mathcal{J}\tau/2} \\
0%
\end{array}%
\right){\normalcolor .}
\end{equation*}
Since the transformation on the right hand side is an $SU(4)$
transformation, the solution in Eq.~(\ref{eq:partemb}) is equivalent to $%
z^{1}=z_{3}^{\dagger}=\frac{1}{\sqrt{2}}e^{i\mathcal{J}\tau/2}$, $%
z^{2}=z^{4}=0$. Furthermore, the engineering
dimension of this operator is $J$, which matches the energy of the point-particle
solution, and the two-loop dilatation operator in Eq.~(\ref{eq:dilat})
vanishes when applied to this operator, which is consistent with our
expectation that the anomalous dimension of the operator should vanish.

\subsection{Excitation Spectrum}
We summarize the spectrum of fluctuations obtained with the algebraic curve
and the world sheet in Table~\ref{tabBMN}. The algebraic curve frequencies have been re-scaled by a factor of $\kappa$ in order to compare them to the world-sheet frequencies. The derivations of these
frequencies are described in Appendices D and E. Note that the fluctuations
in this table are labeled by polarizations. Although this notation was only
defined for the algebraic curve formalism, we find that the world-sheet
frequencies match the algebraic curve frequencies up to constant shifts, so
it is convenient to label the world-sheet frequencies with polarizations as
well. Also note that both sets of frequencies agree with the spectrum of
fluctuations that were found in the Penrose limit (up to constant shifts) \cite{Nishioka:2008gz, Grignani:2008is, Gaiotto:2008cg}.
\begin{table}[tb]
\caption{Spectrum of fluctuations about the point-particle solution computed
using the world-sheet (WS) and algebraic curve (AC) formalisms ($\protect\omega_n=\protect\sqrt{n^2+\protect\kappa^2}$).
Polarizations (\textcolor{red}{\bf{heavy}}/\textcolor{blue}{\bf{light}}) indicate which
pairs of sheets are connected by a fluctuation in the AC formalism, and $%
\pm$ indicates that half of the frequencies have a + and the other half have
a $-$. }
\label{tabBMN}%
\begin{equation}
\begin{array}{c|c|c|l}
\toprule & \mathrm{\mathbf{WS}} & \mathrm{\mathbf{AC}} & \mathrm{\mathbf{%
Polarizations}} \\
\midrule {\rm \bf AdS} &
\begin{array}{c}
\omega_n%
\end{array}
&
\begin{array}{c}
\omega_n%
\end{array}
&
\begin{array}{c}
\textcolor{red}{\bf{(1,10);(2,9);(1,9)}}%
\end{array}
\\
\midrule {\rm \bf Fermions} &
\begin{array}{c}
\omega_n\pm\frac{\kappa}{2} \\
\frac{1}{2}\omega_{2n}%
\end{array}
&
\begin{array}{c}
\omega_n-\frac{\kappa}{2} \\
\frac{1}{2}\omega_{2n}%
\end{array}
&
\begin{array}{l}
\textcolor{red}{\bf{(1,7);(1,8);(2,7);(2,8)}} \\
\textcolor{blue}{\bf{(1,5);(1,6);(2,5);(2,6)}}%
\end{array}
\\
\midrule CP^{3} &
\begin{array}{c}
\omega_n \\
\frac{1}{2}\omega_{2n}\pm\frac{\kappa}{2}%
\end{array}
&
\begin{array}{c}
\omega_n-\kappa \\
\frac{1}{2}\omega_{2n}-\frac{\kappa}{2}%
\end{array}
&
\begin{array}{l}
\textcolor{red}{\bf{(3,7)}} \\
\textcolor{blue}{\bf{(3,5);(3,6);(4,5);(4,6)}}%
\end{array}
\\
\bottomrule
\end{array}
\notag
\end{equation}%
\end{table}

While the constant shifts in the world-sheet spectrum occur with opposite
signs and can be removed by gauge transformations, this is not
the case for the algebraic curve frequencies. In fact, the constant shifts
in the algebraic curve frequencies have physical significance, which can be
seen by taking the mode number $n=0$. In this limit, the $AdS$ frequencies reduce to~$\kappa$, the $%
CP^{3}$ frequencies reduce to 0, and the fermi frequencies reduce to~$\kappa/2$. In
this sense, the $n=0$ algebraic curve frequencies have {}``flat-space'' behavior. This property was also observed for algebraic curve frequencies computed about solutions in $AdS_{5} \times S^{5}$ \cite{Gromov:2007aq}.
On the other hand, the world-sheet frequencies do not have this property. In
the next subsection, we will see that the constant shifts in the algebraic
curve spectrum have important implications for the one-loop correction to
the classical energy.

\subsection{One-Loop Correction to Energy}
Using Eqs.~(\ref{eq:light}) and (\ref{eq:heavy}) we see that $\omega_{H}$
and $\omega_{L}$ are constants for both the world-sheet and algebraic curve
spectra. In particular, for the world-sheet spectrum we find that $%
\omega_{H}(n)=\omega_{L}(n)=0$. As a result, both the standard summation
prescription in Eq.~(\ref{eq:badsum}) and the new summation prescription in
Eq.~(\ref{eq:goodsum}) give a vanishing one-loop correction to the energy.
On the other hand, for the algebraic curve we find that $\omega_{H}(n)=\kappa$
and $\omega_{L}(n)=-2\kappa$. For these values of $\omega_{H}$ and $\omega_{L}$,
the new summation prescription gives a vanishing one-loop correction but the
standard summation prescription gives a linear divergence:
\begin{equation*}
\delta E_{1-loop,old}=\lim_{N\rightarrow\infty}-(N+1/2){\normalcolor .}
\end{equation*}
Thus we find that both summation prescriptions are consistent with
supersymmetry if we use the spectrum computed from the world-sheet, but only
the new summation is consistent with supersymmetry if we use the spectrum
computed from the algebraic curve.

\section{Spinning String}

\subsection{Classical Solution and Dual Operator}
In the global coordinates of Eqs.~(\ref{Ads4metric}) and (\ref{cp3}), the
solution for a circular spinning string with two equal nonzero spins in $%
CP^{3}$ is
\begin{equation}
t=\kappa \tau ,\,\,\,\rho =0,\,\,\,\xi =\pi /4,\,\,\,\theta _{1}=\theta
_{2}=\pi /2,\,\,\,\psi =m\sigma ,\,\,\,\phi _{1}=\phi _{2}=2\mathcal{J}\tau
,  \label{eq:spinglob}
\end{equation}%
where $\mathcal{J}=\frac{J}{4\pi g}$ and $m$ is the winding number. Using Eqs.~(\ref{eq:adsmap}) and (\ref{eq:mapping}), we can also
write this solution in embedding coordinates (which are useful for doing
algebraic curve calculations):
\begin{equation}
n_{1}=\cos \kappa \tau ,\,\,\,n_{2}=\sin \kappa \tau
,\,\,\,n_{3}=n_{4}=n_{5}=0,\,\,\,z^{1}=z_{4}^{\dagger }=\frac{1}{2}e^{i\left( \mathcal{J}\tau +m\sigma/2
\right) },\,\,\,z^{3}=z_{2}^{\dagger }=\frac{1}{2}e^{i\left( \mathcal{J}\tau
-m\sigma/2 \right) }.  \label{eq:spinemb}
\end{equation}%
Equations (\ref{eq:E}-\ref{Js}) imply that $E=4\pi g\kappa $, $S=0$, $J_{\psi }=0$, and $J_{\phi _{1}}=J_{\phi _{2}}=J$. Furthermore, the Virasoro constraints
in Eq.~(\ref{eq:Virasoro}) give $\kappa =\sqrt{m^{2}+4\mathcal{J}^{2}}$, or
equivalently $E=2J\sqrt{1+\frac{\pi ^{2}m^{2}\lambda }{2J^{2}}}$. In the
limit $\mathcal{J}\gg m$, this reduces to the BPS condition $E=2J$, so we
expect that the dual operator should have engineering dimension $2J$ and a
finite but non-zero anomalous dimension. Furthermore, the dispersion
relation has a BMN expansion in the parameter $\lambda /J^{2}$, which allows
us to make a prediction for anomalous dimension of the dual operator.
Expanding the dispersion relation to first order in the BMN parameter gives
\begin{equation}
E=2J+\frac{\pi ^{2}m^{2}\lambda }{2J}+\mathcal{O}\left( \lambda
^{2}/J^{3}\right) {\normalcolor.}  \label{eq:circexpand}
\end{equation}%
To extrapolate this formula to the gauge theory, we must make the
replacement $\lambda \rightarrow 2\lambda ^{2}$. One way to understand this
replacement is by comparing the magnon dispersion relation at strong and
weak t'Hooft coupling, as explained in the introduction. We therefore get
the following prediction for the anomalous dimension of the dual gauge
theory operator%
\begin{equation}
{\bf \Delta} -2J=\frac{\pi ^{2}\lambda ^{2}m^2}{J}+\mathcal{O}\left( \lambda^2 /J^{2}\right) .
\label{eq:spinad}
\end{equation}%
The higher order terms in the expansion of the classical
string energy in Eq.~(\ref{eq:circexpand}) correspond to $\mathcal{O}\left(
\lambda ^{4}/J^{3}\right) $ corrections to the anomalous dimension, but the
one-loop correction to the energy provides $\mathcal{O}%
\left( \lambda ^{2}/J^{2}\right) $ corrections to the anomalous dimension (see Eq.~\ref{prediction}).

The dual gauge theory operator should have the form
\begin{equation}
\mathcal{O}={\rm tr}\left[\left( Z^{1}Z_{2}^{\dagger }\right) ^{J}\left(
Z^{3}Z_{4}^{\dagger }\right) ^{J}+...\right],
\label{dualop}\end{equation}
where the dots stand for permutations of $\left( Z^{1}Z_{2}^{\dagger
}\right) $ and $\left( Z^{3}Z_{4}^{\dagger }\right) $. Note that the engineering dimension of the
operator is $2J$, as expected. When we apply the two loop dilatation
operator in Eq.~(\ref{eq:dilat}) to the operator in Eq.~(\ref{dualop}), it reduces to
\begin{equation}
{\bf \Delta} -2J=\lambda ^{2}\sum_{i=1}^{2J}\left(
1-P_{2i-1,2i+1}+1-P_{2i,2i+2}\right) {\normalcolor.}  \label{eq:cp3su2}
\end{equation}%
This is the Hamiltonian for two identical Heisenberg spin chains; one
located on the even sites and the other on the odd sites. If one thinks of $Z^{1}$
and $Z_{2}^{\dagger }$ as being up spins and $Z^{3}$ and
$Z_{4}^{\dagger }$ as being down spins, then each spin chain has $J$ up
spins and $J$ down spins. In appendix H, we use the Bethe ansatz to show that the anomalous dimension is indeed given by Eq.~(\ref{eq:spinad}).

\subsection{Excitation Spectrum}
We summarize the spectrum of fluctuations about the spinning string in Tables~\ref{tabAC}~and~\ref{freqT}. The algebraic curve frequencies have been re-scaled by a factor of $\kappa$ in order to compare them to the world-sheet frequencies. The derivations are presented in Appendices F and G.
\begin{table}[tbp]
\caption{Spectrum of fluctuations about the spinning string solution
computed using the world-sheet (WS) and algebraic curve (AC) formalisms. The notation for the frequencies is given in Table~\protect\ref%
{freqT}. Polarizations (\textcolor{red}{\bf{heavy}}/\textcolor{blue}{\bf{light}})
indicate which pairs of sheets are connected by a fluctuation in the AC
formalism. }
\label{tabAC}%
\begin{equation}
\begin{array}{c|c|c|l}
\toprule & \mathrm{\mathbf{WS}} & \mathrm{\mathbf{AC}} & \mathrm{\mathbf{%
Polarizations}} \\
\midrule {\rm \bf AdS} &
\begin{array}{l}
\omega_n^{A}%
\end{array}
&
\begin{array}{l}
\omega_n^{A}%
\end{array}
&
\begin{array}{l}
\textcolor{red}{\bf{(1,10);(2,9);(1,9)}}%
\end{array}
\\
\midrule {\rm \bf Fermions} &
\begin{array}{l}
\omega_n^{F}+\frac{\kappa}{2} \\
\omega_n^{F}-\frac{\kappa}{2} \\
\omega_{2n}^{A}/2%
\end{array}
&
\begin{array}{l}
\omega_n^{F}+\frac{\kappa}{2}-2\mathcal{J} \\
\omega_{m+n}^{F}-\frac{\kappa}{2} \\
\omega_{2n}^{A}/2%
\end{array}
&
\begin{array}{l}
\textcolor{red}{\bf{(1,7);(2,7)}} \\
\textcolor{red}{\bf{(1,8);(2,8)}} \\
\textcolor{blue}{\bf{(1,5);(1,6);(2,5);(2,6)}}%
\end{array}
\\
\midrule CP^{3} &
\begin{array}{l}
\omega_n^{C} \\
\omega_n^{C_{-}} \\
\omega_n^{C_{+}}%
\end{array}
&
\begin{array}{l}
\omega_{m+n}^{C}-2\mathcal{J} \\
\omega_{m+n}^{C_{-}} \\
\omega_n^{C_{+}}-2\mathcal{J}%
\end{array}
&
\begin{array}{l}
\textcolor{red}{\bf{(3,7)}} \\
\textcolor{blue}{\bf{(3,5);(3,6)}} \\
\textcolor{blue}{\bf{(4,5);(4,6)}}%
\end{array}
\\
\bottomrule
\end{array}
\notag
\end{equation}%
\end{table}
\begin{table}[tb]
\caption{Notation for spinning string frequencies.}
\label{freqT}%
\begin{equation}
\begin{array}{l|l}
\toprule \mathrm{\mathbf{\quad\quad\quad eigenmodes}} & \mathrm{\mathbf{%
notation}} \\
\midrule
\begin{array}{l}
\sqrt{2\mathcal{J}^2+n^2\pm \sqrt{4\mathcal{J}^4+n^2\kappa^2}} \\
\sqrt{4\mathcal{J}^2+n^2-m^2} \\
\end{array}
&
\begin{array}{l}
\omega_n^{C_\pm} \\
\omega_n^{C} \\
\end{array}
\\
\midrule \sqrt{n^2+\kappa^2} &
\begin{array}{l}
\omega_n^{A}%
\end{array}
\\
\midrule \sqrt{4\mathcal{J}^2+n^2} &
\begin{array}{l}
\omega_n^{F}%
\end{array}
\\
\bottomrule
\end{array}
\notag
\end{equation}%
\end{table}
We find
that the algebraic curve spectrum matches the world-sheet spectrum up to
constant shifts and shifts in mode number. Furthermore, if we set the
winding number $m=0$ and take $\mathcal{J}\rightarrow\mathcal{J}/2$, we find
that all the frequencies in Table~\ref{tabAC} reduce to the
corresponding frequencies in Table~\ref{tabBMN}, which is expected since
setting the winding number to zero reduces the string to a point-particle.\footnote{In showing that the (3,5), (3,6), (4,5), and (4,6) frequencies in Table~\ref{tabAC} reduce to those in Table~\ref{tabBMN}, the identity in  Eq.~(\ref%
{sqrtident}) is useful.}
This is an important check of our results for the spinning string.
On the other hand, if we set the mode number $n=0$, we find that the
algebraic curve frequencies once again have flat-space behavior, i.e., the $AdS$
frequencies reduce to~$\kappa$, the $CP^{3}$ frequencies reduce to 0, and the fermi
frequencies reduce to~$\kappa/2$.

Finally, we would like to point out that both the
algebraic curve and world-sheet spectra have instabilities when $|m|\geq2$. For example if we set $m=2$, then the algebraic curve frequencies
labeled by (3,5) and (3,6) become imaginary for $n=-3$ and $n=-1$ and the
corresponding world-sheet frequencies become imaginary for $n=\pm1$. \footnote{We would like to thank Victor Mikhaylov for showing us his unpublished notes on the spinning string algebraic curve \cite{notes}. In these notes, he also derives the algebraic curve for the spinning string and uses it to compute the fluctuation frequencies, however the asymptotics that he imposes on the algebraic curve are different than the asymptotics we use in Eq.~(\ref{eq:asymptotics}). The differences occur in the signs of several terms on the right hand side of Eq.~(\ref{eq:asymptotics}). As a result, we obtain frequencies with different constant shifts.}

\subsection{One-Loop Correction to the Energy}
For the spinning string, $\omega _{H}(n)$ and $\omega _{L}(n)$ defined in
Eqs.~(\ref{eq:light}) and (\ref{eq:heavy}) are nontrivial:
\begin{eqnarray*}
\omega _{H}^{WS}(n) &=&3\omega _{n}^{A}+\omega _{n}^{C}-4\omega _{n}^{F},\,\,
\\
\,\omega _{L}^{WS}(n) &=&2\omega _{n}^{C_{+}}+2\omega _{n}^{C_{-}}-2\omega
_{2n}^{A}, \\
\omega _{H}^{AC}(n) &=&3\omega _{n}^{A}+\omega _{n+m}^{C}-2\omega
_{n}^{F}-2\omega _{n+m}^{F}+2\mathcal{J},\,\, \\
\omega _{L}^{AC}(n) &=&2\omega _{n}^{C_{+}}+2\omega _{n+m}^{C_{-}}-2\omega
_{2n}^{A}-4\mathcal{J},
\end{eqnarray*}%
where WS stands for world-sheet, AC stands for algebraic curve, and we used the notation in Table~\ref{freqT}.

To compute the one-loop correction, we must evaluate an infinite
sum of the form
\begin{equation}
\delta E_{1-loop}=\sum_{n=-\infty}^{\infty}\Omega\left(\mathcal{J},n,m\right){\normalcolor .}
\label{sum1}
\end{equation}
Note that the frequency $\Omega$ in this equation should not be confused with the off-shell frequencies defined in section 2. Since we have two summation prescriptions (the old one in Eq.~(\ref{eq:badsum}) and the new one in Eq.~(\ref{eq:goodsum})) and two sets of frequencies (world-sheet and algebraic curve) there are four choices for $\Omega\left(\mathcal{J},n,m\right)$:
\begin{equation}
\begin{array}{l}
\Omega _{old,WS} =\frac{1}{2\kappa}\left(\omega_{H}^{WS}(n)+\omega_{L}^{WS}(n)\right),\,\,\, \\
\Omega _{new,WS} =\frac{1}{2\kappa}\left(2\omega_{H}^{WS}(2n)+\omega_{L}^{WS}(n)\right), \\
\Omega _{old,AC} =\frac{1}{2\kappa}\left(\omega_{H}^{AC}(n)+\omega_{L}^{AC}(n)\right), \\
\Omega _{new,AC} =\frac{1}{2\kappa}\left(2\omega_{H}^{AC}(2n)+\omega_{L}^{AC}(n)\right),
\end{array}
\label{summands}
\end{equation}
where $old$/$new$ refers to the summation prescription.

To gain further insight, let's look at the summands in Eq.~(\ref{summands}) in two limits: the large-$n$ limit and the large-$\mathcal{J}$ limit. By looking at the large-$n$ limit, we will learn about the convergence properties of the one-loop corrections, and by looking at the large-$\mathcal{J}$ limit and evaluating the sums over $n$ using $\zeta$-function regularization, we will be able to compute the $\mathcal{J}^{-2n}$ contributions to the one-loop corrections. These are referred to as the analytic terms. In general there can also be terms proportional to $\mathcal{J}^{-2n+1}$, which are referred to as the non-analytic terms, and exponentially suppressed terms, i.e. terms that scale like $e^{-\mathcal{J}}$. These terms are sub-dominant compared to the analytic terms in the large-$\mathcal{J}$ limit.

\subsubsection*{Large-n limit}

Note that in all four cases $\Omega\left(\mathcal{J},-n,m\right)=\Omega\left(\mathcal{J},n,-m\right)$, so the one-loop correction in Eq.~(\ref{sum1}) can be written as
\begin{equation}
\delta E_{1-loop}=\Omega\left(\mathcal{J},0,m\right)+\sum_{n=1}^{\infty}\left(\Omega\left(\mathcal{J},n,m\right)+\Omega\left(\mathcal{J},n,-m\right)\right)
\label{sum2}.
\end{equation}
The large-$n$ limit of $\Omega\left(\mathcal{J},n,m\right)+\Omega\left(\mathcal{J},n,-m\right)$ for the four choices of $\Omega\left(\mathcal{J},n,m\right)$ is summarized in Table~ \ref{largen}.
\begin{table}[tb]
\caption{Large-$n$ limit of  $\Omega\left(\mathcal{J},n,m\right)+\Omega\left(\mathcal{J},n,-m\right)$ for the old summation (where $\Omega\left(\mathcal{J},n,m\right)=\omega_{H}(n)+\omega_{L}(n)$) and the new summation (where $\Omega\left(\mathcal{J},n,m\right)=2\omega_{H}(2n)+\omega_{L}(n)$) applied to the world-sheet (WS) spectrum and algebraic curve (AC) spectrum.}
\label{largen}%
\begin{equation}
\begin{array}{c|c|c}
\toprule & \mathrm{\mathbf{WS}} & \mathrm{\mathbf{AC}} \\
\midrule {\mathbf{Old\,\,Sum}} &
\begin{array}{l}
-\frac{m^{2}\left(5m^{2}/4+3\mathcal{J}^{2}\right)}{2\kappa n^{3}}+\mathcal{O}\left(n^{-5}\right)
\end{array}
&
\begin{array}{l}
-\frac{2\mathcal{J}}{\kappa}-\frac{m^{2}\left(11m^{2}/4+5\mathcal{J}^{2}\right)}{2\kappa n^{3}}+\mathcal{O}\left(n^{-4}\right)
\end{array}
\\
\midrule {\mathbf{New\,\,Sum}} &
\begin{array}{l}
-\frac{m^{4}}{4\kappa n^{3}}+\mathcal{O}\left(n^{-5}\right)
\end{array}
&
\begin{array}{l}
\frac{m^{2}\left(\mathcal{J}^{2}-5m^{2}/4\right)}{2\kappa n^{3}}+\mathcal{O}\left(n^{-4}\right)
\end{array}
\\
\bottomrule
\end{array}
\notag
\end{equation}%
\end{table}

From this table we see that all one-loop
corrections are free of quadratic and logarithmic divergences because terms of order $n$ and order $1/n$ cancel out in
the large-$n$ limit. At the same time, we find a linear divergence when we apply the old summation prescription to the algebraic curve spectrum since the summand has a constant term. In all other cases however, the summands are
at most $\mathcal{O}(n^{-3})$, which suggests that the one-loop corrections are
convergent. Hence we find that both summation prescriptions give finite one-loop corrections when applied to the world-sheet spectrum, but only the new summation prescription gives a finite result when applied to the algebraic
curve spectrum. This is the same thing we found for the point-particle. The
new feature of the spinning string is that the one-loop correction is
nonzero and therefore provides a nontrivial prediction to be compared with the dual gauge theory.

\subsubsection*{Large-$\mathcal{J}$ limit}
In the previous section we found that when $\Omega\left(\mathcal{J},n,m\right)=\Omega _{old,AC}\left(\mathcal{J},n,m\right)$, the one-loop correction is divergent but for the other three cases in Eq.~(\ref{summands}), it is convergent. This means we have three possible predictions for the one-loop correction, however by expanding the summands in the large-$\mathcal{J}$ limit and evaluating the sums over $n$ at each order of $\mathcal{J}$ using $\zeta$-function regularization, we find that all three cases give the same result. The technique of $\zeta$-function
regularization is convenient for computing the analytic terms in the
large-$\mathcal{J}$ expansion of one-loop corrections but does not capture non-analytic and exponentially suppressed terms \cite{SchaferNameki:2005tn}. We now describe this procedure in
more detail.

If we expand in the summand in the large-$\mathcal{J}$ limit, only even powers of $\mathcal{J}$ appear:
\begin{equation}
\sum_{n=-\infty}^{\infty}\Omega\left(\mathcal{J},n,m\right)=\sum_{k=1}^{\infty}\mathcal{J}^{-2k}\sum_{n=-\infty}^{\infty}\Omega_{k}(n,m){\normalcolor .}\label{eq:j1}\end{equation}
For each power of $\mathcal{J}$, the sum over $n$ can be written
as follows
\begin{equation}
\sum_{n=-\infty}^{\infty}\Omega_{k}(n,m)=\Omega_{k}(0,m)+\sum_{n=1}^{\infty}\left(\Omega_{k}(n,m)+\Omega_{k}(n,-m)\right){\normalcolor .}\label{eq:j2}\end{equation}
If we expand $\Omega_{k}(n,m)$ in the limit $n\rightarrow\infty$,
we find that it splits into two pieces:

\[
\Omega_{k}(n,m)=\sum_{j=-1}^{2k}c_{k,j}(m)n^{j}+\tilde{\Omega}_{k}(n,m)\]
where $\tilde{\Omega}_{k}(n,m)$ is $\mathcal{O}\left(n^{-2}\right)$. We will refer to $\tilde{\Omega}_{k}(n,m)$
as the finite piece because it converges when summed over $n$, and
$\sum_{j=-1}^{2k}c_{k,j}(m)n^{j}$ as the divergent piece because
it diverges when summed over $n$. Furthermore, we find that $\tilde{\Omega}_{k}(n,m)=\tilde{\Omega}_{k}(n,-m)$
and $c_{k,j}(m)\propto m^{2k-j}$. Hence, the odd powers of $n$ cancel
out of the divergent piece when we add $\Omega_{k}(n,m)$ to $\Omega_{k}(n,-m)$
and we get \[
\Omega_{k}(n,m)+\Omega_{k}(n,-m)=2\left[\sum_{j=0}^{k}c_{k,2j}(m)n^{2j}+\tilde{\Omega}_{k}(n,m)\right]{\normalcolor .}\]
Noting that $\zeta(0)=-1/2$ and $\zeta\left(2j\right)=0$ for $j>0$,
we see that only the constant term in the divergent piece contributes
if we evaluate the sum over $n$ using $\zeta$-function regularization:
\[
\sum_{n=1}^{\infty}\left(\Omega_{k}(n,m)+\Omega_{k}(n,-m)\right)\rightarrow-c_{k,0}+2\sum_{n=1}^{\infty}\tilde{\Omega}_{k}(n,m){\normalcolor .}\]
Combining this with Eqs.~(\ref{eq:j1}) and (\ref{eq:j2}) then gives

\[
\delta E_{1-loop}=\sum_{k=1}^{\infty}\mathcal{J}^{-2k}\left[\Omega_{k}(0,m)-c_{k,0}+2\sum_{n=1}^{\infty}\tilde{\Omega}_{k}(n,m)\right]{\normalcolor .}\]

Using the procedure described above, we obtain a single prediction
for the one-loop correction to the energy of the spinning string:
\begin{eqnarray}
\delta E_{1-loop} &=&\frac{1}{2\mathcal{J}^{2}}\left[ m^{2}/4+\sum_{n=1}^{%
\infty }\left( n\left( \sqrt{n^{2}-m^{2}}-n\right) +m^{2}/2\right) \right]
\label{eq:predicold} \\
&&-\frac{1}{8\mathcal{J}^{4}}\left[ 3m^{4}/16+\sum_{n=1}^{\infty }\left(
\begin{array}{c}
3m^{4}/8-n^{4} \\
+n\sqrt{n^{2}-m^{2}}\left( m^{2}/2+n^{2}\right)  \notag
\end{array}%
\right) \right] +\mathcal{O}\left( \frac{1}{\mathcal{J}^{6}}\right){%
\normalcolor. \notag}
\end{eqnarray}%
In showing that Eq.~(\ref{sum1}) gives this prediction when $\Omega(\mathcal{J},n,m)=\Omega _{new,AC}(\mathcal{J},n,m)$, it is convenient to shift the index of summation as follows: $\Omega_{new,AC}(\mathcal{J},n,m)\rightarrow\Omega_{new,AC}(\mathcal{J},n-m,m)$.
Since the sum is convergent, this shift does not change its value.
Recalling that $\mathcal{J}=\frac{J}{\sqrt{2\pi^{2}\lambda}}$ and making the replacement $\lambda\rightarrow2\lambda^{2}$ in Eq.~(\ref{eq:predicold}) then gives a prediction for the $1/J$ correction to the anomalous
dimension of the gauge theory operator in Eq. (\ref{dualop}):
\begin{equation}
\Delta-2J=\left(\frac{\pi^{2}\lambda^{2}m^2}{J}+...\right)+\frac{1}{J}\left(\frac{2a\pi^{2}\lambda^{2}}{J}+...\right)\label{prediction}\end{equation}
where
\begin{eqnarray*}
a &=&m^2/4+\sum_{n=1}^{\infty }\left( n\left( \sqrt{n^{2}-m^2}-n\right) +m^2/2\right) . \\
\end{eqnarray*}%
Note that the first term in Eq.~(\ref{prediction}) came from expanding the classical dispersion relation for the spinning string to first order in the BMN parameter $\lambda/J^2$ and then making the replacement $\lambda\rightarrow2\lambda^{2}$. We verify Eq.~(\ref{prediction}) from the gauge theory side in Appendix H.

\section{Conclusions}

In this paper, we study various methods for computing one-loop corrections to the energies of  classical solutions to type IIA string theory in $AdS_{4} \times \CP$. Previous studies which computed the one-loop correction to the folded spinning string in $AdS_{4}$ found that agreement with the all-loop $AdS_{4}/CFT_{3}$ Bethe ansatz is not achieved using the standard summation prescription that was used for type IIB string theory in $AdS_{5} \times S^{5}$. Rather, a new summation prescription seems to be required which distinguishes between so-called light modes and heavy modes. We extend this investigation by analyzing the one-loop correction to the energy of a point-particle and a circular spinning string, both of which are located at the spatial origin of $AdS_{4}$ and have nontrivial support in $\CP$. The spinning string considered in this paper has two equal nonzero spins in $CP^{3}$ and is
the analogue of the $SU(2)$ spinning string in $AdS_{5}\times S^{5}$. The point-particle and spinning string are important examples to analyze because they have trivial support in $AdS_{4}$ and therefore avoid the $\kappa$-symmetry
issues that arise for solutions which purely have support in $AdS_{4}$, such as the folded spinning string.

We use two techniques to compute the spectrum of fluctuations about these solutions. One
technique, called the world-sheet approach, involves expanding the GS action to quadratic order in the fluctuations and computing the normal modes of the resulting action. The other technique, called the algebraic curve approach, involves computing the algebraic curve
for the classical solutions and then carrying out semi-classical
quantization. For the point-particle, we find that the world-sheet and algebraic curve fluctuation frequencies match the spectrum of fluctuations obtained in the Penrose limit up to constant shifts. Furthermore, for the spinning string we find that the algebraic curve spectrum matches the world-sheet spectrum up to constant shifts and shifts in mode number. In particular, the AC and WS frequencies for the spinning string both reduce to the corresponding point-particle frequencies when the winding number is set to zero and become unstable when the winding number $|m|\geq2$. This is familiar from the $SU(2)$ spinning string in $AdS_{5}\times S^{5}$ \cite{Frolov:2003qc,Tseytlin:2003ii} which has instabilities for $|m|\geq1$.

Although the algebraic curve spectrum looks very similar to the world-sheet spectrum, it exhibits some important differences. For example, we find that the algebraic curve frequencies have flat-space behavior when the mode number $n=0$. This was also found for algebraic curve frequencies in $AdS_{5} \times S^{5}$. More importantly, if we compute one-loop corrections by adding up the algebraic curve frequencies using the standard summation prescription that was used in $AdS_5 \times S^{5}$, then we get a linear divergence. This is inconsistent with supersymmetry because we expect the one-loop correction to vanish for the point-particle and to be nonzero but finite for the spinning string. We propose a new summation prescription in Eq. (\ref{eq:goodsum}) which gives precisely these results when applied both to the algebraic curve spectrum and the world-sheet spectrum. This summation prescription has certain similarities to the one that was proposed in
\cite{Gromov:2008fy}. In particular, it also gives a one-loop correction to the folded spinning string which agrees with the all-loop Bethe ansatz. At the same time, it has some important differences which are described in section 2.3. For example, we find that our summation prescription generally gives more well-defined results for one-loop corrections.

In principle we can get three predictions for the one-loop correction to the spinning string (one coming from the algebraic curve and two coming from the world-sheet since the world-sheet gives finite results using both the old summation prescription in Eq. (\ref{eq:badsum}) and the new summation prescription in Eq. (\ref{eq:goodsum})), but by expanding the one-loop corrections in the large-$\mathcal{J}$ limit (where $\mathcal{J}=\frac{J}{\sqrt{2\lambda}\pi}$ and $J$ is the spin) and evaluating the sum at each order in $\mathcal{J}$ using $\zeta$-function regularization, we find that all three cases actually give the same result. This is very nontrivial considering that our new summation prescription looks very different than the old one. Furthermore, we show that this result agrees with the predictions of the Bethe ansatz. Thus, while the old summation prescription only seems to work when applied to the world-sheet frequencies of solutions with trivial support in $AdS_{4}$, our summation prescription works more generally. Fully understanding why the old summation prescription breaks down for solutions with nontrivial support in $AdS_{4}$ warrants further study.

It would be useful to confirm our results using methods more rigorous than $\zeta$-function regularization. This can be done using the contour integral techniques developed in~\cite{SchaferNameki:2006gk}, which can also be used to compute $1/\mathcal{J}^{2n+1}$ and exponentially suppressed terms in the large-$\mathcal{J}$ expansion of the one-loop corrections. It would also be interesting to evaluate the one-loop correction to the spinning string energy in a way that does not rely on summation prescriptions. The basic idea would be to identify the one-loop correction with a
normal ordering constant which can be then determined by demanding that the
quantum generators of certain symmetries preserved by the classical solution
have the right algebra. Something along these lines was done for the type
IIB superstring in plane-wave background in~\cite{Metsaev:2002re}. Ultimately, fully understanding how to compute one-loop corrections to type IIA string theory in $AdS_4 \times \CP$ may lead to important tests of the $AdS_{4}/CFT_{3}$ correspondence.

\section*{Acknowledgments}

This work was supported in part by the U.S. Department of Energy under Grant No. DE-FG02-92ER40701.
We are grateful to Shinji Hirano, Tristan McLoughlin, Victor Mikhaylov, Tatsuma Nishioka, Sakura Schafer-Nameki, and John H.
Schwarz for helpful discussions. In particular, we would like to thank VM
for sharing his unpublished notes on the spinning string algebraic curve
and its semi-classical quantization, SSN for
helping us with various calculations in this paper and for providing many
useful explanations, and JHS for his guidance and
many useful comments.

\appendix
\appendixpage
\addappheadtotoc

\section{Review of ABJM}
The ABJM theory is a three-dimensional superconformal Chern-Simons gauge theory with $\mathcal{N}=6$ supersymmetry. The bosonic field content consists of four complex scalars $Z^{1}$,$Z^{2}$, $Z^{3}$, $Z^{4}$ and their adjoints $Z^{\dagger}_{1}$,$Z^{\dagger}_{2}$, $Z^{\dagger}_{3}$, $Z^{\dagger}_{4}$ (which transform in the $(\bar{N},N)$ and $(N,\bar{N})$ representations of the gauge group $U(N)\times U(N)$) as well as two $U(N)$ gauge fields $A_{\mu}$ and $\hat{A}_{\mu}$. The kinetic and Chern-Simons terms for these fields are
\[\mathcal{L}_{kin}=-\frac{k}{2\pi}{\rm tr}\left(D_{\mu}Z^{I}D^{\mu}Z_{I}^{\dagger}\right)\]
\[\mathcal{L}_{CS}=\frac{k}{2\pi}\epsilon^{\mu\nu\lambda}{\rm tr}\left(\frac{1}{2}A_{\mu}\partial_{\nu}A_{\lambda}+\frac{i}{3}A_{\mu}A_{\nu}A_{\lambda}-\frac{1}{2}\hat{A}_{\mu}\partial_{\nu}\hat{A}_{\lambda}-\frac{i}{3}\hat{A}_{\mu}\hat{A}_{\nu}\hat{A}_{\lambda}\right),\]
where $D_{\mu}Z^{I}=\partial_{\mu}Z^{I}+i\left(A_{\mu}Z^{I}-Z^{I}\hat{A}_{\mu}\right)$ and $k$ is called the level. For the complete action see \cite{Bandres:2008ry, Benna:2008zy}. The scalars have mass dimension $1/2$ and transform in the fundamental representation of the R-symmetry group $SU(4)$. Their adjoints transform in the anti-fundamental representation
of $SU(4)$. The theory has a large-$N$ expansion with 't Hooft parameter $\lambda=N/k$. For $k=1,2$, the theory is conjectured to have $\mathcal{N}=8$ supersymmetry.  For $k\ll N\ll k^{5}$, the theory is conjectured to be dual to type IIA string theory on $AdS_{4}\times \CP$.

For operators of the form
\[
\mathcal{O}=W_{k_{1}...k_{J}}^{i_{1}...i_{J}}{\rm tr}\left(Z^{k_{1}}Z_{i_{1}}^{\dagger}...Z^{k_{J}}Z_{i_{J}}^{\dagger}\right)\]
the two-loop dilatation operator is given by
\begin{equation}
{\bf \Delta}-J=\frac{\lambda^{2}}{2}\sum_{i=1}^{2J}\left(2-2P_{i,i+2}+P_{i,i+2}T_{i,i+1}+T_{i,i+1}P_{i,i+2}\right)\label{eq:dilat}\end{equation}
where $\lambda=N/k$, $P$ is the permutation operator, and $T$ is the trace operator \cite{Minahan:2008hf}. Note that the indices are periodic, i.e. $2J+1 \sim 1$ and $2J+2 \sim 2$.


\section{$AdS_{4}\times CP^{3}$ Geometry}
We use $M,N=$ $\left( \mathbf{0},\mathbf{1},\ldots \mathbf{9}\right) $ to
label base-space indices and $A,B=\left( 0,1,\ldots 9\right) $ to label
tangent-space indices. We assign the first four indices to $AdS_{4}$
and the last six indices to $CP^{3}$. In this appendix, we take the $AdS_{4}$
and $CP^{3}$ spaces to have unit radii. A radius $R$ can be readily
incorporated by $ds^{2}\rightarrow R^{2}ds^{2}$ and $e_{M}{}^{A}\rightarrow
Re_{M}{}^{A}.$

\subsection{$AdS_{4}$}

The metric for an $AdS_{4}$ space with unit radius in global coordinates $%
(t,\rho ,\theta ,\phi )$ is given by%
\begin{equation}
ds^{2}_{AdS_4}=-\cosh ^{2}\rho \mathrm{d}t^{2}+\mathrm{d}\rho ^{2}+\sinh ^{2}\rho
\left( \mathrm{d}\theta ^{2}+\sin ^{2}\theta \mathrm{d}\phi ^{2}\right) ,
\end{equation}%
where $-\infty <t<\infty ,$ $0\leq \rho <\infty ,$ $0\leq \theta \leq\pi ,$ $%
0\leq \phi <2\pi .$

The embedding coordinates are defined by%
\begin{equation}
n_{1}^{2}+n_{2}^{2}-n_{3}^{2}-n_{4}^{2}-n_{5}^{2}=1,  \label{emb1}
\end{equation}%
and they are related to the global coordinates by%
\begin{equation}
\begin{array}{l}
n_{1}=\cosh \rho \cos t, \\
n_{2}=\cosh \rho \sin t, \\
n_{3}=\sinh \rho \cos \theta \sin \phi , \\
n_{4}=\sinh \rho \sin \theta \sin \phi , \\
n_{5}=\sinh \rho \cos \phi .%
\end{array}
\label{eq:adsmap}
\end{equation}%
Because the global coordinates are not well defined at $\rho =0,$ it is
useful to define Cartesian coordinates $\left( t,\eta _{1},\eta _{2},\eta
_{3}\right)={\bf(0,1,2,3)}$ for which the metric is given by%
\begin{equation}
ds_{AdS_{4}}^{2}=g_{MN}^{AdS_{4}}dX^{M}dX^{N}=\frac{1}{\left( 1-\eta
^{2}\right) ^{2}}\left[ -\left( 1+\eta ^{2}\right) ^{2}dt^{2}+4d\vec{\eta}%
\cdot d\vec{\eta}\right] {\normalcolor.}  \label{eq:ads4}
\end{equation}%
Note that this metric is only valid for $\eta ^{2}=\vec{\eta}\cdot \vec{\eta}%
=\eta _{1}^{2}+\eta _{2}^{2}+\eta _{3}^{2}<1$. These coordinates are related to the global coordinates by $\cosh \rho
=(1+\eta ^{2})/(1-\eta ^{2})$.

The vielbein (defined by $g_{MN}^{AdS_{4}}=e_{M}{}^{A}e_{N}{}^{B}\eta_{AB}$ where $%
\mathbf{\eta }_{AB}=diag\left( -1,1,1,1\right)$) is given by
\begin{equation*}
e_{M}^{}{A}=\left(
\begin{array}{cccc}
\frac{\left( 1+\eta ^{2}\right) }{\left( 1-\eta ^{2}\right) } & 0 & 0 & 0 \\
0 & \frac{2}{\left( 1-\eta ^{2}\right) } & 0 & 0 \\
0 & 0 & \frac{2}{\left( 1-\eta ^{2}\right) } & 0 \\
0 & 0 & 0 & \frac{2}{\left( 1-\eta ^{2}\right) }%
\end{array}%
\right) ,\bigskip
\end{equation*}%
where $M={\bf (0,1,2,3)}$ labels the rows and $A=(0,1,2,3)$ labels the columns.

The nonzero components of the spin connection $\left( \omega
_{M}{}{}^{AB}=-\omega _{M}{}{}^{BA}{}\right) $ are%
\begin{eqnarray*}
\omega _{\mathbf{0}}{}^{01} &=&2\eta _{1}/\left( 1-\eta ^{2}\right) ,\quad
\omega _{\mathbf{0}}{}{}{}^{02}=2\eta _{2}/\left( 1-\eta ^{2}\right) ,\quad
\omega _{\mathbf{0}{}}{}{}^{03}=-2\eta _{3}/\left( 1-\eta ^{2}\right) ,\quad
\\
\omega _{\mathbf{1{}}}{}{}^{12} &=&2\eta _{2}/\left( 1-\eta ^{2}\right)
,\quad \omega _{\mathbf{1}}{}{}^{13}=2\eta _{3}/\left( 1-\eta ^{2}\right)
,\quad  \\
\omega _{\mathbf{2}}{}{}^{21} &=&2\eta _{1}/\left( 1-\eta ^{2}\right) ,\quad
\omega _{\mathbf{2}}{}{}^{23}=2\eta _{3}/\left( 1-\eta ^{2}\right) ,\quad  \\
\omega _{\mathbf{3}}{}{}^{31} &=&2\eta _{1}/\left( 1-\eta ^{2}\right) ,\quad
\omega _{\mathbf{3}}{}{}^{32}=2\eta _{2}/\left( 1-\eta ^{2}\right) .\quad
\end{eqnarray*}

\subsection{$CP^{3}$}

The metric for a $CP^{3}$ space with unit radius in global coordinates $%
\left( \psi ,\xi ,\varphi _{1},\theta _{1},\varphi _{2},\theta _{2}\right)
={\bf (4,5,6,7,8,9)}$ is given by%
\begin{eqnarray}
ds_{\CP}^{2} &=&g_{MN}^{CP^{3}}dX^{M}dX^{N}=\mathrm{d}\xi ^{2}+\cos ^{2}\xi \sin
^{2}\xi \left( \mathrm{d}\psi +\frac{1}{2}\cos \theta _{1}\mathrm{d}\varphi
_{1}-\frac{1}{2}\cos \theta _{2}\mathrm{d}\varphi _{2}\right) ^{2}
\label{cp32} \\
&&+\frac{1}{4}\cos ^{2}\xi \left( \mathrm{d}\theta _{1}^{2}+\sin ^{2}\theta
_{1}\mathrm{d}\varphi _{1}^{2}\right) +\frac{1}{4}\sin ^{2}\xi \left(
\mathrm{d}\theta _{2}^{2}+\sin ^{2}\theta _{2}\mathrm{d}\varphi
_{2}^{2}\right) ,  \notag
\end{eqnarray}%
where $0\leq \xi <\pi /2,$ $0 \leq \psi < 2\pi ,$ $0\leq \theta _{i}\leq \pi ,$ and $0\leq \varphi _{i}<2\pi $ \cite{Cvetic:2000yp,Nishioka:2008gz,Bergman:2009zh}. The $CP^{3}$ Kähler form is given by $\bf{J}$$=dA$
where%
\begin{equation}
A=\frac{1}{2}\left( \cos \theta _{1}\cos ^{2}\xi d\phi _{1}+\cos \theta
_{2}\sin ^{2}\xi d\phi _{2}+\cos 2\xi d\psi \right) {.}  \label{eq:kalerform}
\end{equation}

The embedding or homogeneous coordinates ($z^{I}\in {\mathbb{C}}$) are defined by%
\begin{equation}
\sum_{I=1}^{4}\left\vert z^{I}\right\vert ^{2}=1,\qquad z^{I}\sim
e^{i\lambda }z^{I},  \label{emb2}
\end{equation}%
where $\lambda \in {\mathbb{R}}$. The embedding coordinates are related to the global coordinates by
\begin{equation}
\begin{array}{l}
z_{1}=\cos \xi \cos \frac{\theta _{1}}{2}\exp \left( i\frac{\psi +\varphi
_{1}}{2}\right) , \\
z_{2}=\cos \xi \sin \frac{\theta _{1}}{2}\exp \left( i\frac{\psi -\varphi
_{1}}{2}\right) , \\
z_{3}=\sin \xi \cos \frac{\theta _{2}}{2}\exp \left( i\frac{-\psi +\varphi
_{2}}{2}\right) , \\
z_{4}=\sin \xi \sin \frac{\theta _{2}}{2}\exp \left( i\frac{-\psi -\varphi
_{2}}{2}\right) .%
\end{array}
\label{eq:mapping}
\end{equation}%
Note that the metric in Eq. \ref{cp32} can be written in terms of embedding coordinates as follows:
\[
ds_{CP^{3}}^{2}= dz\cdot dz^{\dagger }-\left( z^{\dagger }\cdot dz \right)
\left( z \cdot dz^{\dagger } \right)
\]
where $z \cdot z^{\dagger}=\sum_{I=1}^{4} z^{I} z^{\dagger}_{I}$.

The vielbein (defined by $g_{MN}^{CP^{3}}=e_{M}{}^{A}e_{N}{}^{B}\delta _{AB}$) is
\begin{equation*}
e_{M}{}^{A}=\left(
\begin{array}{cccccc}
\cos \xi \sin \xi  & 0 & 0 & 0 & 0 & 0 \\
0 & 1 & 0 & 0 & 0 & 0 \\
\cos \xi \sin \xi \cos \theta _{1}/2 & 0 & \cos \xi \sin \theta _{1}/2 & 0 &
0 & 0 \\
0 & 0 & 0 & \cos \xi /2 & 0 & 0 \\
-\cos \xi \sin \xi \cos \theta _{2}/2 & 0 & 0 & 0 & \sin \xi \sin \theta
_{2}/2 & 0 \\
0 & 0 & 0 & 0 & 0 & \sin \xi /2%
\end{array}%
\right) ,
\end{equation*}%
where $M={\bf (4,5,6,7,8,9)}$ labels the rows and $A=(4,5,6,7,8,9)$ labels the columns.

The nonzero components of the spin connection $\left( \omega
_{M}{}{}^{AB}=-\omega _{M}{}{}^{BA}{}\right) $ are
\begin{equation*}
\begin{array}{lll}
\omega _{\mathbf{4}}{}{}^{45}=\cos \left( 2\xi \right) , & \omega _{\mathbf{4%
}}{}{}^{76}=\sin ^{2}\xi , & \omega _{\mathbf{4{}}}{}^{89}=\cos ^{2}\xi , \\
\omega _{\mathbf{6{}}}{}^{45}=\cos \theta _{1}\cos \left( 2\xi \right) /2, &
\omega _{\mathbf{6{}}}{}{}^{74}=\omega _{\mathbf{6}{}}{}^{56}=\sin \theta
_{1}\sin \xi /2, &  \\
\omega _{\mathbf{6{}}}{}^{67}=-\cos \theta _{1}(\sin ^{2}\xi -2)/2, & \omega
_{\mathbf{6{}}}{}^{89}=\cos \theta _{1}\cos ^{2}\xi /2, &  \\
\omega _{\mathbf{7{}}}{}^{46}=\omega _{\mathbf{7{}}}{}^{57}{}=\sin \xi /2, &
&  \\
\omega _{\mathbf{8}}{}^{54}=\cos \theta _{2}\cos \left( 2\xi \right) /2, &
\omega _{\mathbf{8}}{}^{49}=\omega _{\mathbf{8}}{}^{85}=\sin \theta _{2}\cos
\xi /2, &  \\
\omega _{\mathbf{8}}{}^{67}=\cos \theta _{2}\sin ^{2}\xi /2, & \omega _{%
\mathbf{8}}{}^{98}=\cos \theta _{2}(\cos ^{2}\xi -2)/2, &  \\
\omega _{\mathbf{9}}{}{}^{84}=\omega _{\mathbf{9}}{}^{95}=\cos \xi /2. &  &
\end{array}%
\end{equation*}%

\subsection{Fluxes}

Using the vielbein, we can convert between base-space and
tangent-space coordinates. In particular, by writing the four-form field
strength in Eq.~(\ref{F4}) in tangent-space coordinates, one finds that
\begin{equation}
F_{ABCD}=\frac{6k}{R^{2}}\epsilon _{ABCD},  \label{eq:4form}
\end{equation}%
where $\epsilon _{0123}=1$ and all other non-zero components are related by
antisymmetry. Furthermore, if one takes the exterior derivative of Eq.~(\ref%
{eq:kalerform}), plugs this into Eq. (\ref{F2}), and converts to tangent-space coordinates, one finds that
\begin{equation}
F_{AB}=\frac{2k}{R^{2}}\epsilon _{AB},  \label{eq:2form}
\end{equation}%
where $\epsilon _{45}=\epsilon _{67}=\epsilon _{89}=1$ and all other
non-zero components are related by anti-symmetry. Equations (\ref{Dilaton}),
(\ref{eq:4form}), and (\ref{eq:2form}) then imply that
\begin{eqnarray*}
e^{\phi }\Gamma \cdot F_{2} &=&\frac{2}{R}\left( \Gamma ^{45}+\Gamma
^{67}+\Gamma ^{89}\right) ,\,   \\
\,\,e^{\phi }\Gamma \cdot F_{4} &=&\frac{6}{R}\Gamma ^{0123}{.}
\end{eqnarray*}
Plugging these expressions into Eq. \ref{GammaF} then gives
\begin{equation}
\Gamma\cdot F=\frac{1}{4R}\left[-\Gamma_{11}\left(\Gamma^{45}+\Gamma^{67}+\Gamma^{89}\right)+3\Gamma^{0123}\right].
\label{GammaF2}
\end{equation}


\section{Dirac Matrices}

We use the following representation of the 10d Dirac matrices ($%
\left\{ \Gamma ^{A},\Gamma ^{B}\right\} =2\eta ^{AB}$):%
\begin{equation*}
\begin{array}{lll}
\Gamma ^{0}=i\gamma ^{0}\otimes \mathrm{I}\otimes \mathrm{I}\otimes \mathrm{I%
},\quad  & \Gamma ^{1}=i\gamma ^{1}\otimes \mathrm{I}\otimes \mathrm{I}%
\otimes \mathrm{I}\quad  & \Gamma ^{2}=i\gamma ^{2}\otimes \mathrm{I}\otimes
\mathrm{I}\otimes \mathrm{I},\quad  \\
\Gamma ^{3}=i\gamma ^{3}\otimes \mathrm{I}\otimes \mathrm{I}\otimes \mathrm{I%
}, & \Gamma ^{4}=\gamma ^{5}\otimes \sigma _{2}\otimes \mathrm{I}\otimes
\sigma _{1}, & \Gamma ^{5}=\gamma ^{5}\otimes \sigma _{2}\otimes \mathrm{I}%
\otimes \sigma _{3}, \\
\Gamma ^{6}=\gamma ^{5}\otimes \sigma _{1}\otimes \sigma _{2}\otimes \mathrm{%
I}, & \Gamma ^{7}=\gamma ^{5}\otimes \sigma _{3}\otimes \sigma _{2}\otimes
\mathrm{I}, & \Gamma ^{8}=\gamma ^{5}\otimes \mathrm{I}\otimes \sigma
_{1}\otimes \sigma _{2}, \\
\Gamma ^{9}=\gamma ^{5}\otimes \mathrm{I}\otimes \sigma _{3}\otimes \sigma
_{2}, &  &
\end{array}%
\end{equation*}%
where $\mathrm{I}$ is the $2\times 2$ identity matrix, the $\gamma ^{\prime
}s$ are 4d Dirac matrices given by%
\begin{equation*}
\begin{array}{ll}
\gamma ^{0}=\sigma _{1}\otimes \mathrm{I},\qquad  & \gamma ^{1}=i\sigma
_{2}\otimes \sigma _{1}, \\
\gamma ^{2}=i\sigma _{2}\otimes \sigma _{2},\qquad  & \gamma ^{3}=i\sigma
_{2}\otimes \sigma _{3}, \\
\gamma ^{5}=i\gamma ^{0}\gamma ^{1}\gamma ^{2}\gamma ^{3}, &
\end{array}%
\end{equation*}%
and the Pauli matrices are given by
\begin{equation*}
\sigma _{1}=\left[
\begin{array}{cc}
0 & 1 \\
1 & 0%
\end{array}%
\right] ,\qquad \sigma _{2}=\left[
\begin{array}{cc}
0 & -i \\
i & 0%
\end{array}%
\right] ,\qquad \sigma _{3}=\left[
\begin{array}{cc}
1 & 0 \\
0 & -1%
\end{array}%
\right] .
\end{equation*}%
Finally, we define the 10d chirality operator as%
\begin{equation*}
\Gamma_{11}=\Gamma ^{0}\Gamma ^{1}\Gamma ^{2}\Gamma ^{3}\Gamma ^{4}\Gamma
^{5}\Gamma ^{6}\Gamma ^{7}\Gamma ^{8}\Gamma ^{9}.
\end{equation*}



\section{Point-Particle Spectrum from the World-Sheet}

\subsection{Bosonic Spectrum}

To compute the spectrum of bosonic fluctuations about the point-particle,
first add fluctuations to the classical solution in Eq.~(\ref{eq:partglob}):%
\begin{equation*}
t=\kappa \tau +\delta t(\tau ,\sigma ),\,\,\,\eta _{i}=\delta \eta _{i}(\tau ,\sigma ),\,\,\,\xi=\pi
/4+\delta \xi (\tau ,\sigma ),
\end{equation*}%
\begin{equation*}
\theta _{j}=\pi /2+\delta \theta _{j}(\tau ,\sigma
),\,\,\,\psi=\kappa \tau +\delta \psi (\tau ,\sigma
),\,\,\,\phi _{j}=\delta \phi _{j}(\tau ,\sigma ),
\end{equation*}%
where $i=1,2,3$ and $j=1,2$. Expanding the bosonic Lagrangian in Eq.~(\ref%
{eq:bosaction}) to quadratic order gives%
\begin{eqnarray*}
4\pi \mathcal{L}_{bos} &=&-\frac{1}{4}\left( \partial \delta t\right) ^{2}+%
\frac{1}{4}\left( \partial \delta \psi \right) ^{2}+\sum_{i=1}^{3}\left[
\left( \partial \delta \eta _{i}\right) ^{2}+\kappa ^{2}\delta \eta _{i}^{2}%
\right] +\left( \partial \delta \xi \right) ^{2}+\kappa ^{2}\delta \xi ^{2}
\\
&&+\frac{1}{8}\left( \partial \delta \theta _{1}\right) ^{2}+\frac{1}{8}%
\left( \partial \delta \theta _{2}\right) ^{2}+\frac{1}{8}\left( \partial
\delta \phi _{1}\right) ^{2}+\frac{1}{8}\left( \partial \delta \phi
_{2}\right) ^{2}+\frac{1}{4}\kappa \delta \theta _{1}\delta \dot{\phi}_{1}-%
\frac{1}{4}\kappa \delta \theta _{2}\delta \dot{\phi}_{2},
\end{eqnarray*}%
where $\left( \partial f\right) ^{2}=-\left( \partial _{\tau }f\right)
^{2}+\left( \partial _{\sigma }f\right) ^{2}$. We immediately see that the
fluctuations $\delta t$ and $\delta \psi $ are massless, while $\delta \eta
_{i}$ and $\delta \xi $ have mass $\kappa $. If we consider Fourier modes of the form $f(\tau ,\sigma )=\tilde{f}e^{i(\omega \tau +n\sigma )}$,
then the equations of motion for the remaining fields reduce to
\begin{equation*}
\left(
\begin{array}{cccc}
\omega ^{2}-n^{2} & -i\omega \kappa  & 0 & 0 \\
i\omega \kappa  & \omega ^{2}-n^{2} & 0 & 0 \\
0 & 0 & \omega ^{2}-n^{2} & i\omega \kappa  \\
0 & 0 & -i\omega \kappa  & \omega ^{2}-n^{2}%
\end{array}%
\right) \left(
\begin{array}{c}
\delta \tilde{\theta}_{1} \\
\delta \tilde{\phi}_{1} \\
\delta \tilde{\theta}_{2} \\
\delta \tilde{\phi}_{2}%
\end{array}%
\right) =0{\normalcolor.}
\end{equation*}%
The dispersion relations for the normal modes of this system are obtained by
taking the determinant of the matrix on the left-hand side, setting it to
zero, and solving for $\omega $. The positive solutions are%
\begin{equation}
\frac{\omega }{\kappa }=\sqrt{\frac{1}{4}+\frac{n^{2}}{\kappa ^{2}}}\pm
\frac{1}{2}{\normalcolor.}  \label{eq:bmnlight}
\end{equation}%
Each of these solutions has multiplicity two, giving a total of four
positive solutions.

In summary, we find that there are eight massive modes and two massless
modes. Three of the massive modes come from $AdS_{4}$. Their dispersion
relations are given by
\begin{equation*}
\frac{\omega}{\kappa}=\sqrt{1+\frac{n^{2}}{\kappa^{2}}}{\normalcolor .}
\end{equation*}
The remaining five massive modes come from $CP^{3}$. One of them has the
dispersion relation in the equation above and the other four have the
dispersion relations in Eq.~(\ref{eq:bmnlight}). The two massless modes are
longitudinal and can be discarded. This can be seen by expanding the Virasoro
constraints in Eq.~(\ref{eq:Virasoro}) to linear order in the perturbations.
Doing so gives
\begin{equation*}
\partial_{\tau}\left(\delta
t-\delta\psi\right)=\partial_{\sigma}\left(\delta t-\delta\psi\right)=0{%
\normalcolor .}
\end{equation*}
Noting that $\left(\partial\delta
t\right)^{2}-\left(\partial\delta\psi\right)^{2}=\partial\left(\delta
t-\delta\psi\right)\partial\left(\delta t+\delta\psi\right)$, we see that
the equation above implies that all terms in the action involving $\delta t$
and $\delta\psi$ vanish.

\subsection{Fermionic Spectrum}
In order to compute the spectrum of fermionic fluctuations
about the point-particle solution given by Eq. (\ref{eq:partglob}), we only need to know the pullback of the vielbein and the
spin connection in the background of this classical solution. These are given by
\begin{equation}
\label{e1}
e_{\tau} =\frac{R}{2}\mathcal{J}\left( -\Gamma ^{0}+\Gamma ^{4}\right),\,\,\,\,\,\,\,\ e_{\sigma} =0,
\end{equation}
and%
\begin{equation}
\label{w1}
\omega _{\tau} =\mathcal{J}\left( \Gamma ^{89}-\Gamma ^{67}\right),\,\,\,\,\,\,\,\ \omega _{\sigma} =0.
\end{equation}
Plugging these expressions into Eq. (\ref{Proj}) gives
\begin{equation}
P_{+}= \frac{1}{2}\left( 1+\Gamma ^{0}\Gamma ^{4}\right) ,
\label{ProjB}
\end{equation}%
where we used $K=\partial _{\tau}X^{\mu }e_{\mu }^{0}=(R/2)\mathcal{J}$.
It is straightforward to check that Eqs.~(\ref{ct1},\ref{ct2}) are satisfied for the point-particle solution.
Therefore, by plugging Eqs.~(\ref{GammaF2},\ref{w1},\ref{ProjB}) into Eq.~(\ref{Feom}) and using the Dirac matrices in appendix B, we obtain
an explicit form of the equation of motion for the fermionic fluctuations.
The frequencies are then determined using the procedure described at the end of section 2.1. In particular, the positive fermionic frequencies are given by%
\begin{eqnarray*}
\omega _{1} &=&\sqrt{\kappa ^{2}+n^{2}}, \\
\omega _{2} &=&\frac{1}{2}\sqrt{\kappa ^{2}+4n^{2}}+\frac{\kappa }{2}, \\
\omega _{3} &=&\frac{1}{2}\sqrt{\kappa ^{2}+4n^{2}}-\frac{\kappa }{2},
\end{eqnarray*}%
where $\omega _{2}$ and $\omega _{3}$ have multiplicity two while $\omega
_{1}$ has multiplicity four, for a total of 8 fermionic frequencies.


\section{Point-Particle Algebraic Curve}

\subsection{Classical Quasimomenta}

\begin{figure}[tb]
\center
\includegraphics{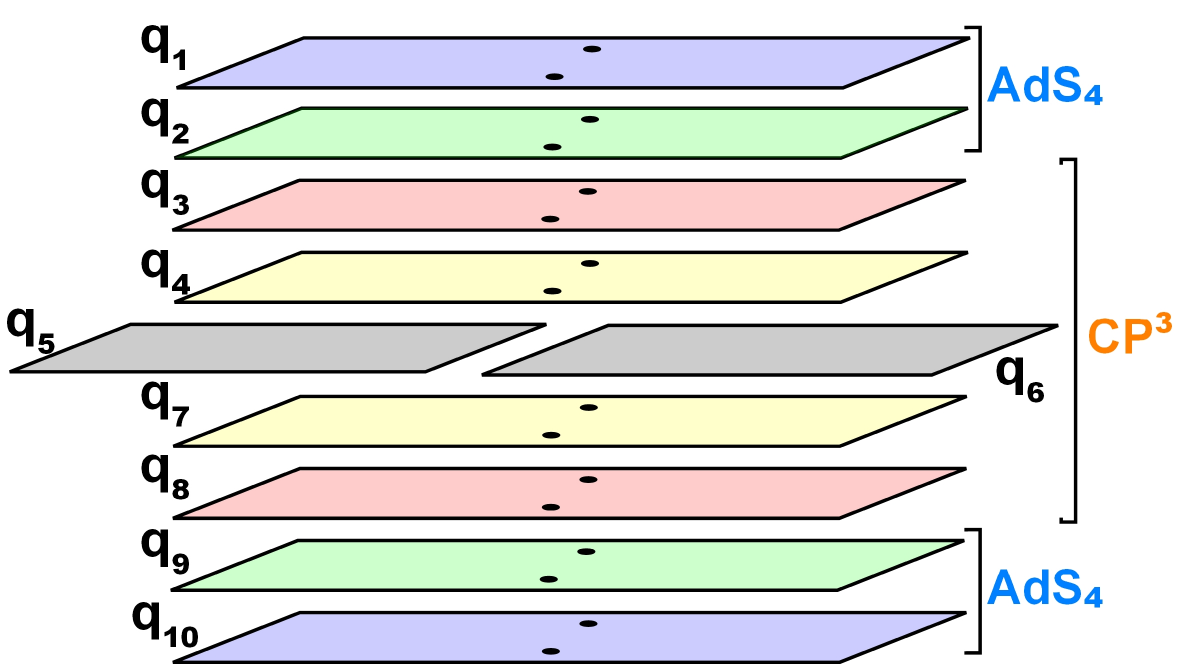}
\caption{Classical algebraic curve for the point-particle rotating in $\CP$.}
\label{BMN}
\end{figure}

In this section, we compute the algebraic curve for the classical
solution given in Eq.~(\ref{eq:partemb}). First we plug this solution into Eq.~(\ref{eq:curr}):
\begin{equation*}
\left(j_{\tau}\right)_{AdS_{4}}=2\kappa\left(%
\begin{array}{ccccc}
0 & 1 & 0 & 0 & 0 \\
-1 & 0 & 0 & 0 & 0 \\
0 & 0 & 0 & 0 & 0 \\
0 & 0 & 0 & 0 & 0 \\
0 & 0 & 0 & 0 & 0%
\end{array}%
\right),\,\,\,\left(j_{\tau}\right)_{CP^{3}}=i\mathcal{J}\left(%
\begin{array}{cccc}
1 & 0 & 0 & 0 \\
0 & 0 & 0 & 0 \\
0 & 0 & -1 & 0 \\
0 & 0 & 0 & 0%
\end{array}%
\right),\,\,\, j_{\sigma}=0{\normalcolor .}
\end{equation*}
Note that this connection is independent of $\sigma$, so it is trivial to
compute the monodromy matrix in Eq.~(\ref{eq:mono}) since path ordering is
not an issue. Diagonalizing the monodromy matrix and comparing the
eigenvalues to Eqs.~(\ref{eq:hat}) and (\ref{eq:tilde}) then gives $\hat{p%
}_{1}=-\hat{p}_{4}=\frac{4\pi\kappa x}{x^{2}-1}$, $\tilde{p}_{1}=-\tilde{p}%
_{4}=\frac{2\pi\mathcal{J}x}{x^{2}-1}$, and $\hat{p}_{2}=\hat{p}_{3}=\tilde{p%
}_{2}=\tilde{p}_{3}=0$. Recalling that $\kappa=\mathcal{J}$ and plugging
these results into Eq.~(\ref{eq:pq}), we find that the classical
quasimomenta are
\begin{equation}
q_{1}=q_{2}=q_{3}=q_{4}=\frac{2\pi\mathcal{J}x}{x^{2}-1},\,\,\, q_{5}=0{%
\normalcolor .}  \label{eq:partquasi}
\end{equation}
The algebraic curve corresponding to these quasimomenta is depicted in Fig.~(%
\ref{BMN}). Note that all sheets except those corresponding to $q_{5}$ and $%
q_{6}$ have poles at $x=\pm1$.

\subsection{Off-shell Frequencies}

Recall from Eqs.~(\ref{LL},\ref{M35}) and table \ref{tLH} that if we know the off-shell
frequencies $\Omega_{15}(y)$ and $\Omega_{45}(y)$, then
all the others are determined. Let's begin by computing $\Omega _{15}(y)$. Suppose we have two fluctuations
between $q_{1}$ and $q_{5}$. To satisfy level-matching, let's take one of
these fluctuations to have mode number $+n$ and the other to have mode
number $-n$. Each fluctuation corresponds to adding a pole to the classical algebraic curve. The locations of the poles are determined by solving
Eq.~\ref{eq:8}. We will denote the pole locations by $x_{\pm n}^{15}$. We then make the following ansatz for the
fluctuations:
\begin{eqnarray*}
\delta q_{1}(x,y) &=&\sum_{\pm }\frac{\alpha \left( x_{n}^{15}\right) }{%
x-x_{n}^{15}},\,\,\,\delta q_{2}(x,y)=-\delta q_{1}(1/x,y),\,\,\, \\
\delta q_{5}(x,y) &=&-\sum_{\pm }\frac{\alpha \left( x\right) }{%
x-x_{n}^{15}}-\sum_{\pm }\frac{\alpha \left(1/x\right) }{%
1/x-x_{n}^{15}},
\end{eqnarray*}%
where $\alpha(x)$ is defined in Eq.~(\ref{eq:alpha}), $\pm $ stands for the sum over the positive and negative mode number, and $y$
is a collective coordinate for the positions of the two poles $x_{\pm
n}^{15} $. We have not made an ansatz for $\delta q_{3}$ and $\delta q_{4}$
because they are not needed to compute $\Omega _{15}(y)$. Notice that this
ansatz satisfies the inversion symmetry in Eq.~(\ref{inversion}) and has pole structure
in agreement with Eq.~(\ref{eq:alpha}). In the large-$x$ limit, the fluctuations reduce
to
\begin{eqnarray*}
\lim_{x\rightarrow \infty }\delta q_{1}(x,y) &\sim &\frac{1}{x}\sum_{\pm
}\alpha \left( x_{n}^{15}\right) ,\,\,\, \\
\lim_{x\rightarrow \infty }\delta q_{2}(x,y) &\sim &\frac{1}{2gx}\sum_{\pm }%
\frac{1}{\left( x_{n}^{15}\right) ^{2}-1},\, \\
\,\,\lim_{x\rightarrow \infty }\delta q_{5}(x,y) &\sim &-\frac{1}{gx},
\end{eqnarray*}%
where we neglect $\mathcal{O}\left( x^{-2}\right) $ terms. Comparing these expressions to Eq.~(\ref{eq:asymptotics})
implies that the anomalous energy shift is given by
\begin{equation*}
\Delta (y)=\sum_{\pm }\frac{1}{\left( x_{n}^{15}\right) ^{2}-1}{\normalcolor.%
}
\end{equation*}%
The off-shell fluctuation frequency is then obtained by plugging this into Eq.~(\ref{eq:9})
and recalling that the (1,5) fluctuation is fermionic:
\begin{equation*}
\Omega _{15}(y)=\Delta (y)+\frac{1}{2}N^{15}=\Delta (y)+1=\sum_{\pm }\frac{1%
}{2}\frac{\left( x_{n}^{15}\right) ^{2}+1}{\left( x_{n}^{15}\right) ^{2}-1}.
\end{equation*}%
This implies that the off-shell frequency for a single fluctuation between $%
q_{1}$ and $q_{5}$ is given by%
\begin{equation*}
\Omega _{15}(y)=\frac{1}{2}\frac{y^{2}+1}{y^{2}-1}{\normalcolor.}
\end{equation*}

Now let's compute $\Omega _{45}(y)$. Once again, let's suppose that we have two
fluctuations between $q_{4}$ and $q_{5}$ which have opposite mode numbers $%
\pm n$. We make the following ansatz for the fluctuations:
\begin{eqnarray*}
\delta q_{1}(x,y) &=&\frac{\alpha _{+}(y)}{x+1}+\frac{\alpha _{-}(y)}{x-1}%
,\,\,\,\delta q_{2}(x,y)=-\delta q_{1}(1/x,y), \\
\delta q_{4}(x,y) &=&-\sum_{\pm }\frac{\alpha \left( x\right) }{x-x_{n}^{45}}%
,\,\,\,\delta q_{3}(x,y)=-\delta q_{4}(1/x,y),\,\,\, \\
\delta q_{5}(x,y) &=&\sum_{\pm }\frac{\alpha \left( x\right) }{x-x_{n}^{45}}%
+\sum_{\pm }\frac{\alpha \left( 1/x\right) }{1/x-x_{n}^{45}},
\end{eqnarray*}%
where $\alpha _{\pm }(y)$ are some functions to be determined. Note that
this ansatz satisfies the inversion symmetry in Eq.~(\ref{inversion}) and has pole structure
in agreement with Eq.~(\ref{eq:alpha}). Taking the large-$x$ limit gives%
\begin{eqnarray*}
\lim_{x\rightarrow \infty }\delta q_{1}(x,y) &\sim &\frac{\alpha
_{+}(y)+\alpha _{-}(y)}{x},\,\,\,\lim_{x\rightarrow \infty }\delta
q_{2}(x,y)\sim \alpha _{-}(y)-\alpha _{+}(y)+\frac{\alpha _{+}(y)+\alpha
_{-}(y)}{x}, \\
\lim_{x\rightarrow \infty }\delta q_{3}(x,y) &\sim
&0,\,\,\,\lim_{x\rightarrow \infty }\delta q_{4}(x,y)\sim -\frac{1}{gx}%
,\,\,\,\lim_{x\rightarrow \infty }\delta q_{5}(x,y)\sim \frac{1}{gx}{%
\normalcolor.}
\end{eqnarray*}%
Comparing these limits with Eq.~(\ref{eq:asymptotics}) implies that%
\begin{equation}
\alpha _{+}(y)=\alpha _{-}(y)=\frac{\Delta (y)}{4g}{\normalcolor.}
\label{eq:apmd}
\end{equation}%
Furthermore, the residues of the poles at $x=\pm 1$ must be synchronized
according to Eq.~(\ref{eq:synch}). For example, if we equate the residues of $\delta
q_{1} $ and $\delta q_{4}$ near $x=+1$ we find that
\begin{equation*}
\lim_{x\rightarrow +1}\delta q_{1}(x,y)\sim \frac{\alpha _{-}(y)}{x-1}%
=\lim_{x\rightarrow +1}\delta q_{4}\sim \left( \frac{1}{4g}\sum_{\pm }\frac{1%
}{x_{n}^{45}-1}\right) \frac{1}{x-1}\rightarrow \alpha _{-}(y)=\frac{1}{4g}%
\sum_{\pm }\frac{1}{x_{n}^{45}-1}{\normalcolor.}
\end{equation*}%
Combining this with the Eq.~(\ref{eq:apmd}) implies that
\begin{equation}
\Delta (y)=\sum_{\pm }\frac{1}{x_{n}^{45}-1}{\normalcolor.}
\label{eq:anom45}
\end{equation}%
At this point it is useful to recall that $x_{n}^{45}$ is a root of the
following equation (which comes from plugging Eq.~(\ref{eq:partquasi})
into Eq.~(\ref{eq:8})):
\begin{equation*}
\frac{2\pi \mathcal{J}x_{n}^{45}}{\left( x_{n}^{45}\right) ^{2}-1}=2\pi n{%
\normalcolor.}
\end{equation*}%
Note that this equation has two roots. The convention that we will follow is
to assign the pole to the root with larger magnitude. Hence, if $n<0$ then $%
x_{n}^{45}=\frac{\mathcal{J}}{n}-\sqrt{1+\frac{\mathcal{J}^{2}}{n^{2}}}$ and
if $n>0$ then $x_{n}^{45}=\frac{\mathcal{J}}{n}+\sqrt{1+\frac{\mathcal{J}^{2}%
}{n^{2}}}$. The point to take away from this discussion is that
\begin{equation*}
x_{+n}^{45}=-x_{-n}^{45}{\normalcolor.}
\end{equation*}%
Using this fact, Eq.~(\ref{eq:anom45}) can be
written as follows:
\begin{equation*}
\Delta (y)=\frac{1}{x_{+n}^{45}-1}-\frac{1}{x_{+n}^{45}+1}=\frac{2}{\left(
x_{+n}^{45}\right) ^{2}-1}=\sum_{\pm }\frac{1}{\left( x_{n}^{45}\right)
^{2}-1}{\normalcolor.}
\end{equation*}%
The off-shell fluctuation frequency is then obtained by plugging this into
Eq.~(\ref{eq:9}) and recalling that the $(4,5)$ fluctuation is a $CP^{3}$ fluctuation:
\begin{equation*}
\Omega _{45}(y)=\Delta (y)=\sum_{\pm }\frac{1}{\left( x_{n}^{45}\right)
^{2}-1}.
\end{equation*}%
It follows that the off-shell frequency for a single fluctuation between $%
q_{4}$ and $q_{5}$ is given by%
\begin{equation*}
\Omega _{45}(y)=\frac{1}{y^{2}-1}{\normalcolor.}
\end{equation*}

The remaining off-shell frequencies are now easily computed from Eqs.~(\ref{LL},\ref{M35}) and table \ref{tLH}.
We summarize the off-shell frequencies in Table~\ref{offshell}.

\begin{table}[tb]
\caption{ Off-shell frequencies for fluctuations about the point-particle
solution. }
\label{offshell}%
\begin{equation}
\begin{array}{c|c|l}
\toprule & \mathrm{\mathbf{\Omega(y)}} & \mathrm{\mathbf{Polarizations}} \\
\midrule {\rm \bf AdS} &
\begin{array}{l}
\frac{y^{2}+1}{y^{2}-1}%
\end{array}
&
\begin{array}{l}
\textcolor{red}{\bf{(1,10);(2,9);(1,9)}}%
\end{array}
\\
\midrule {\rm \bf Fermions} &
\begin{array}{l}
\frac{y^{2}+3}{2(y^{2}-1)} \\
\frac{y^{2}+1}{2(y^{2}-1)} \\
\end{array}
&
\begin{array}{l}
\textcolor{red}{\bf{(1,7);(1,8);(2,7);(2,8)}} \\
\textcolor{blue}{\bf{(1,5);(1,6);(2,5);(2,6)}}%
\end{array}
\\
\midrule CP^{3} &
\begin{array}{l}
\frac{2}{y^{2}-1} \\
\frac{1}{y^{2}-1}%
\end{array}
&
\begin{array}{l}
\textcolor{red}{\bf{(3,7)}} \\
\textcolor{blue}{\bf{(3,5);(3,6);(4,5);(4,6)}}%
\end{array}
\\
\bottomrule
\end{array}
\notag
\end{equation}%
\end{table}

\subsection{On-shell Frequencies}

To compute the on-shell frequencies, we must compute the locations of the
poles by solving Eq.~(\ref{eq:8}). Recall that fluctuations that connect $q_{5}$ or $q_{6}$ to
any other sheets are referred to as light, and all the others are referred
to as heavy. A little
thought shows that for light fluctuations, Eq.~(\ref{eq:8}) reduces to
\begin{equation*}
\frac{\mathcal{J}x_{n}}{x_{n}^{2}-1}=n,
\end{equation*}%
and for heavy fluctuations it reduces to
\begin{equation*}
\frac{\mathcal{J}x_{n}}{x_{n}^{2}-1}=\frac{n}{2}.
\end{equation*}%
Each of these equations admits two solutions. We
will assign the location of the pole to the solution with greater magnitude.
Assuming $n>0$, the location of the pole for light excitations is then
given by
\begin{equation*}
x_{n}=\frac{\mathcal{J}}{2n}+\sqrt{\frac{\mathcal{J}^{2}}{4n^{2}}+1},
\end{equation*}%
and the location of the pole for heavy excitations is given by
\begin{equation*}
x_{n}=\frac{\mathcal{J}}{n}+\sqrt{\frac{\mathcal{J}^{2}}{n^{2}}+1}{%
\normalcolor.}
\end{equation*}%
Plugging these solutions into the off-shell frequencies in Table~\ref{offshell}
readily gives the on-shell algebraic curve frequencies in Table~\ref{tabBMN}.

\section{Spinning String Spectrum from the World-Sheet}

\subsection{Bosonic Spectrum}

In this section we calculate the spectrum of bosonic fluctuations
about the circular spinning string in $AdS_{4} \times CP^{3}$.
Let's begin by adding fluctuations to the solution in Eq.~(\ref{eq:spinglob}):
\begin{eqnarray*}
t&=&\kappa \tau +\delta t(\tau ,\sigma ),\,\,\,\eta_{i}=\delta \eta _{i}(\tau ,\sigma ),\,\,\,\xi=\pi /4+\delta \xi (\tau ,\sigma ), \\
\theta _{j} &=&\pi /2+\delta \theta _{j}(\tau ,\sigma
),\,\,\,\psi=m\sigma +\delta \psi (\tau ,\sigma
),\,\,\,\phi _{j}=2\mathcal{J}\tau +\delta \phi _{j}(\tau
,\sigma ),
\end{eqnarray*}%
where $i=1,2,3$, $j=1,2$, and $\kappa =\sqrt{4\mathcal{J}^{2}+m^{2}}$.
Expanding the bosonic Lagrangian in Eq.~(\ref{eq:bosaction}) to quadratic order
in the fluctuations gives%
\begin{eqnarray*}
4\pi \mathcal{L}_{bos} &=&m^{2}/2-\frac{1}{4}\left( \partial \delta t\right)
^{2}+\sum_{i=1}^{3}\left[ \left( \partial \delta \eta _{i}\right)
^{2}+\kappa ^{2}\delta \eta _{i}^{2}\right] \\
&&+\partial \left( \delta \bar{\psi}\right) ^{2}+\partial \left( \delta \bar{%
\xi}\right) ^{2}-m^{2}\delta \bar{\xi}^{2}+\partial \left( \delta \theta
_{+}\right) ^{2}+4\mathcal{J}^{2}\left( \delta \theta _{+}\right)
^{2}+\partial \left( \delta \theta _{-}\right) ^{2} \\
&&+\partial \left( \delta \phi _{+}\right) ^{2}+\partial \left( \delta \phi
_{-}\right) ^{2}+4\mathcal{J}\left( \delta \theta _{-}\partial _{\tau
}\delta \bar{\psi}+\delta \bar{\xi}\partial _{\tau }\delta \bar{\phi}%
_{-}\right) \\
&&-2m\left( \delta \theta _{-}\partial _{\sigma }\delta \phi
_{+}+\delta \theta _{+}\partial _{\sigma }\delta \phi _{-}\right) ,
\end{eqnarray*}%
where $\delta \bar{\psi}=\sqrt{2}\delta \psi $, $\delta \bar{\xi}=2\sqrt{2}%
\delta \xi $, $\delta \theta _{\pm }=\frac{1}{\sqrt{2}}\left( \delta \theta
_{1}\pm \delta \theta _{2}\right) $, $\delta \phi _{\pm }=\frac{1}{\sqrt{2}}%
\left( \delta \phi _{1}\pm \delta \phi _{2}\right) $, and $\left( \partial
f\right) ^{2}=-\left( \partial _{\tau }f\right) ^{2}+\left( \partial
_{\sigma }f\right) ^{2}$. Note that the $AdS_{4}$ fluctuations are the same
as those of the point-particle. In particular, we see that $\delta t$ is
massless and $\delta \eta _{i}$ have mass $\kappa $. If we consider Fourier modes of the form $f(\tau ,\sigma )=\tilde{f}e^{i(\omega \tau +n\sigma
)} $, then the equations of motion for the $CP^{3}$ fluctuations reduce to
\begin{equation}
\left(
\begin{array}{cccccc}
\omega ^{2}-n^{2}+m^{2} & 2i\mathcal{J}\omega & 0 & 0 & 0 & 0 \\
-2i\mathcal{J}\omega & \omega ^{2}-n^{2} & -imn & 0 & 0 & 0 \\
0 & imn & \omega ^{2}-n^{2}-4\mathcal{J}^{2} & 0 & 0 & 0 \\
0 & 0 & 0 & \omega ^{2}-n^{2} & 0 & -2i\mathcal{J}\omega \\
0 & 0 & 0 & 0 & \omega ^{2}-n^{2} & -imn \\
0 & 0 & 0 & 2i\mathcal{J}\omega & imn & \omega ^{2}-n^{2}%
\end{array}%
\right) \left(
\begin{array}{c}
\delta \widetilde{\bar{\xi}} \\
\delta \widetilde{\phi }_{-} \\
\delta \widetilde{\theta }_{+} \\
\delta \widetilde{\bar{\psi}} \\
\delta \widetilde{\phi }_{+} \\
\delta \widetilde{\theta }_{-}%
\end{array}%
\right) =0.  \label{BosCop}
\end{equation}%
Because the matrix in Eq.~(\ref{BosCop}) is block diagonal, the
fluctuations $\left( \delta \widetilde{\bar{\xi}},\delta \widetilde{\phi }%
_{-},\delta \widetilde{\theta }_{+}\right) $ and $\left( \delta \widetilde{%
\bar{\psi}},\delta \widetilde{\phi }_{+},\delta \widetilde{\theta }%
_{-}\right) $ are decoupled. The frequencies are determined by
taking the determinant of the matrix and finding its roots. The equation we must solve is
\begin{equation*}
\left( n^{2}-\omega ^{2}\right) \left( 4\mathcal{J}^{2}-m^{2}+n^{2}-\omega
^{2}\right) \left( n^{4}-m^{2}n^{2}-\left( 4\mathcal{J}^{2}+2n^{2}\right)
\omega ^{2}+\omega ^{4}\right) ^{2}=0{\normalcolor.}
\end{equation*}%
This polynomial has 12 roots, which come in opposite signs. Of the six
positive roots, three correspond to the fluctuations $\left( \delta
\widetilde{\bar{\xi}},\delta \widetilde{\phi }_{-},\delta \widetilde{\theta }%
_{+}\right) $:
\begin{equation*}
\omega =\sqrt{4\mathcal{J}^{2}+n^{2}-m^{2}},\,\,\,\sqrt{2\mathcal{J}%
^{2}+n^{2}\pm \sqrt{4\mathcal{J}^{4}+n^{2}\kappa ^{2}}},
\end{equation*}%
and three correspond to the fluctuations $\left( \delta \widetilde{\bar{\psi}%
},\delta \widetilde{\phi }_{+},\delta \widetilde{\theta }_{-}\right) $:
\begin{equation*}
\omega =\left\vert n\right\vert ,\,\,\,\sqrt{2\mathcal{J}^{2}+n^{2}\pm \sqrt{%
4\mathcal{J}^{4}+n^{2}\kappa ^{2}}}.
\end{equation*}
Note that the solution $\omega =\left\vert n\right\vert$ corresponds to a massless mode, which can be
discarded along with the other massless mode $\delta t$. The remaining eight
modes are massive and correspond to the transverse degrees of freedom.

\subsection{Fermionic Spectrum}
In this section we compute the spectrum of fermionic fluctuations
about the spinning string solution in Eq. (\ref{eq:spinglob}). The pullback of the vielbein and the
spin connection in the background of this classical solution are given by
\begin{equation}
\label{e1s}
e_{\tau} =R\left( -\frac{\kappa }{2}\Gamma ^{0}+\frac{\mathcal{J}}{\sqrt{2}}\left(\Gamma^{6}+\Gamma^{8}\right)\right), \,\,\,\,\,\ e_{\sigma} = \frac{Rm \Gamma^{4}}{2},
\end{equation}
and%
\begin{equation}
\label{w1s}
\omega _{\tau} =\sqrt{2}\mathcal{J}\left( \Gamma ^{74}+\Gamma ^{85}+\Gamma ^{49}+\Gamma
^{56}\right), \,\,\,\,\,\,\ \omega _{\sigma} =m\left( \Gamma ^{89}+\Gamma ^{76}\right).
\end{equation}
Plugging these expressions into Eq. (\ref{Proj}) then gives
\begin{equation}
P_{+}= \frac{1}{2}\left( 1+\frac{\sqrt{2}\mathcal{J}}{\kappa}\Gamma^{0}\left(\Gamma^{6}+\Gamma^{8}\right)+\frac{m}{\kappa}\Gamma^{0}\Gamma^{4}\Gamma_{11}\right) ,  \label{ProjS}
\end{equation}%
where we used $K=\partial _{\tau}X^{\mu }e_{\mu }^{0}=(R/2)\kappa $.
It is straightforward to check that Eqs.~(\ref{ct1},\ref{ct2}) are satisfied for the spinning string solution.
Therefore, by plugging Eqs.~(\ref{GammaF2},\ref{w1s},\ref{ProjS}) into Eq.~(\ref{Feom}) and using the Dirac matrices in appendix B, we obtain an explicit form of the equation of motion for the fermionic fluctuations.
The frequencies are then determined using the procedure described at the end of section 2.1. In this way, the positive fermionic frequencies are given by%
\begin{eqnarray*}
\omega _{1} &=&\sqrt{4\mathcal{J}^{2}+n^{2}}+\frac{\kappa }{2}, \\
\omega _{2} &=&\sqrt{4\mathcal{J}^{2}+n^{2}}-\frac{\kappa }{2}, \\
\omega _{3} &=&\frac{1}{2}\sqrt{\kappa ^{2}+4n^{2}},
\end{eqnarray*}%
where $\omega _{1}$ and $\omega _{2}$ have multiplicity two while $\omega
_{3}$ has multiplicity four, for a total of 8 fermionic frequencies. In obtaining these expressions, Eq.~\ref{sqrtident} is useful.


\section{Spinning String Algebraic Curve}

\subsection{Classical Quasimomenta}

Since the spinning string has the same motion in $AdS_{4}$ as the point
particle, the $AdS_{4}$ quasimomenta have the same structure and are given by
\begin{equation*}
q_{1}(x)=q_{2}(x)=\frac{2\pi \kappa x}{x^{2}-1},
\end{equation*}%
where $\kappa =\sqrt{4\mathcal{J}^{2}+m^{2}}$ for the spinning string.
Therefore, we just have to find the $CP^{3}$ quasimomenta. For the classical
solution in Eq.~(\ref{eq:spinemb}), one finds that the connection in
Eq.~(\ref{eq:curr}) is given by%
\begin{eqnarray*}
\left( j_{0}\right) _{CP^{3}} &=&i\mathcal{J}\left(
\begin{array}{cccc}
1 & e^{-im\sigma } & 0 & 0 \\
e^{im\sigma } & 1 & 0 & 0 \\
0 & 0 & -1 & -e^{-im\sigma } \\
0 & 0 & -e^{im\sigma } & -1%
\end{array}%
\right) ,\quad \quad \\
\left( j_{1}\right) _{CP^{3}} &=&im\left(
\begin{array}{cccc}
1 & 0 & e^{-2i\mathcal{J}\tau } & 0 \\
0 & -1 & 0 & -e^{-2i\mathcal{J}\tau } \\
e^{2i\mathcal{J}\tau } & 0 & 1 & 0 \\
0 & -e^{2i\mathcal{J}\tau } & 0 & -1%
\end{array}%
\right) .
\end{eqnarray*}%
Using Eq.~(\ref{eq:mono}), the $CP^{3}$ part of the monodromy matrix is
given by
\begin{equation*}
\Lambda (x)=P\exp \frac{1}{x^{2}-1}\int_{0}^{2\pi }d\sigma J(\sigma ,x),
\end{equation*}%
where
\begin{equation}
J(\sigma ,x)=\left(
\begin{array}{cccc}
i\left( \mathcal{J}x+m/2\right) & i\mathcal{J}xe^{-im\sigma } & im/2 & 0 \\
i\mathcal{J}xe^{im\sigma } & i\left( \mathcal{J}x-m/2\right) & 0 & -im/2 \\
im/2 & 0 & -i\left( \mathcal{J}x-m/2\right) & -i\mathcal{J}xe^{-im\sigma } \\
0 & -im/2 & -i\mathcal{J}xe^{im\sigma } & -i\left( \mathcal{J}x+m/2\right)%
\end{array}%
\right) .\quad  \label{eq:lax}
\end{equation}%
and we set $\tau =0$ since the eigenvalues of $\Lambda (x)$ are independent
of $\tau $. At this point, it is useful to observe that under a gauge
transformation of the form $J(\sigma ,x)\rightarrow g^{-1}(\sigma )J(\sigma
,x)g(\sigma )-g^{-1}(\sigma )\partial _{\sigma }g(\sigma )$, the monodromy
matrix transforms as $\Lambda (x)\rightarrow g^{-1}(0)\Lambda \left( x\right)
g(2\pi )$. If $g(0)=\pm g(2\pi )$, then the eigenvalues of $\Lambda (x)$ are
gauge invariant up to a sign. Furthermore, if we can choose $g(\sigma )$ such that the $%
\sigma $-dependence of $J(\sigma ,x)$ is removed, then the monodromy matrix
would be trivial to evaluate since path-ordering would not be an issue. This can be accomplished using the gauge transformation $g\left(\sigma \right) =diag(e^{-im\sigma/2 },e^{im\sigma/2 },e^{-im\sigma/2 },e^{im\sigma/2})$~\cite{notes}. Under this
transformation, $J(\sigma ,x)$ becomes
\begin{equation}
J(\sigma ,x)\rightarrow J(0,x)+i\,\frac{m}{2}\,\,\,diag[1,-1,1,-1]{\normalcolor.}
\label{Jsimp}
\end{equation}
When $m$ is odd, $g(0)=-g(2\pi)$ so we must supplement this gauge transformation with $\Lambda (x)\rightarrow -\Lambda (x)$.

\begin{figure}[tb]
\center
\includegraphics{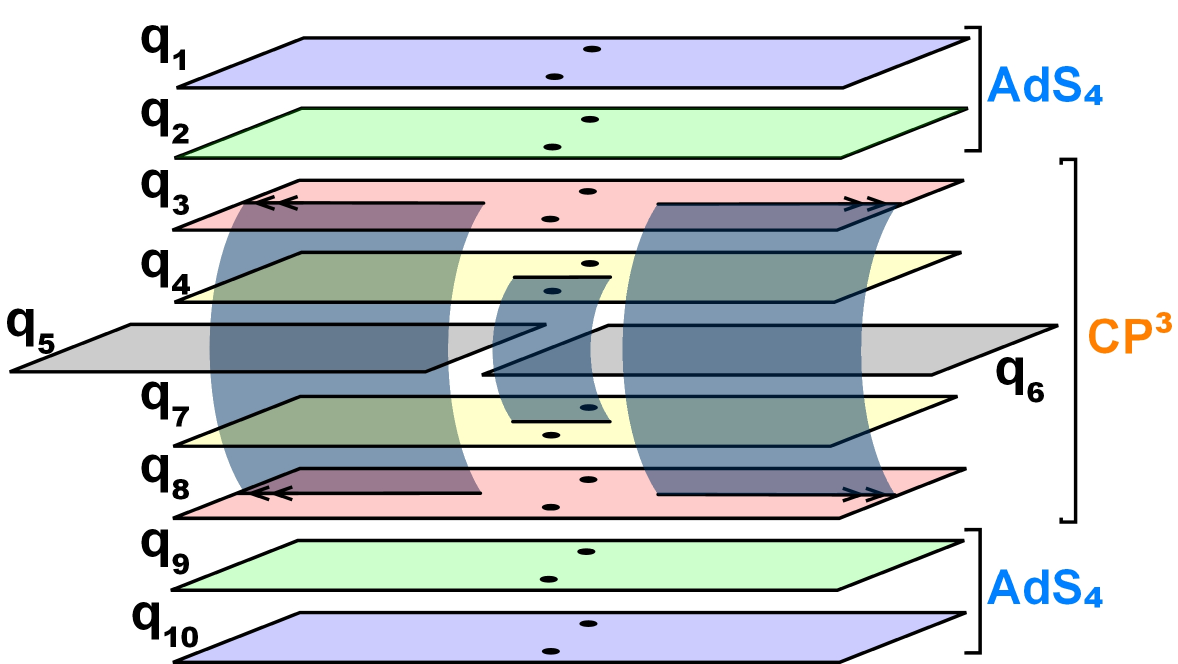}
\caption{Classical algebraic curve for the circular spinning string in $\CP$.}
\label{ACspin}
\end{figure}

Diagonalizing $\Lambda (x)$ and comparing to Eq.~(\ref{eq:tilde}) gives
\begin{equation}  \label{eq:ps}
\begin{array}{l}
\tilde{p}_{1}=\frac{2\pi x}{x^{2}-1}\left[ K(x)+K(1/x)\right]-\pi m , \\
\tilde{p}_{2}=\frac{2\pi x}{x^{2}-1}\left[ K(x)-K(1/x)\right]-\pi m \\
\widetilde{p}_{3}=-\widetilde{p}_{2},\qquad \widetilde{p}_{4}=-\widetilde{p}_{1}, \\
\end{array}%
\end{equation}
where $K(x)=\sqrt{\mathcal{J}^{2}+m^{2}x^{2}/4}$. In deriving Eq. (\ref{eq:ps}), we made use of the following identity:
\begin{equation}
\sqrt{A\pm\sqrt{B}}=\frac{1}{2}\left(\sqrt{2A+2\sqrt{A^{2}-B}}\pm\sqrt{2A-2%
\sqrt{A^{2}-B}}\right){\normalcolor .}  \label{sqrtident}
\end{equation}
Furthermore, we subtracted $\pi m$ from $\tilde{p}_{1}$ and $\tilde{p}_{2}$ and added $\pi m$ to $\tilde{p}_{3}$
and $\tilde{p}_{4}$ so that the quasimomenta are $\mathcal{O}(1/x)$ in
the large-$x$ limit. This also implements the transformation $\Lambda (x)\rightarrow -\Lambda (x)$ when $m$ is odd.
The quasimomenta $q_{3}(x)$, $q_{4}(x)$, and $q_{5}(x)$
are then given by plugging Eq.~(\ref{eq:ps}) into Eq.~(\ref{eq:pq})
\begin{equation}  \label{spinquasi}
\begin{array}{l}
q_{3}(x) =\frac{4\pi x}{x^{2}-1}K(x)-2\pi m, \\
q_{4}(x) =-q_{3}(1/x)-2\pi m=\frac{4\pi x}{x^{2}-1}K(1/x), \\
q_{5}(x) =0.%
\end{array}%
\end{equation}
From these quasimomenta, we see that the spinning string algebraic curve has
a cut between $q_{3}$ and $q_{8}$ and between $q_{4}$ and $q_{7}$ (by inversion symmetry). The classical algebraic curve is depicted
in Fig.~(\ref{ACspin}).

\subsection{Off-shell Frequencies}

Since $q_{1}$ and $q_{5}$ have the same structure as they did for
the point-particle solution, a little thought shows that $\Omega_{15}(y)$ should be the same as we found for the point-particle.
In particular,
\begin{equation*}
\Omega_{15}(y)=\frac{1}{2}\frac{y^{2}+1}{y^{2}-1}{%
\normalcolor .}
\end{equation*}
From Eqs.~(\ref{LL},\ref{M35}) and table \ref{tLH}, it follows that the only off-shell frequency
we need to compute is $\Omega_{45}(x)$.

Let's suppose that we have two fluctuations between $q_{4}$ and $q_{5}$;
one with mode number $+n$ and the other with mode number $-n$. These fluctuations correspond to adding poles to the classical algebraic curve. The locations
of the poles will be denoted $x_{\pm
n}^{45}$. Looking at Eq.~(\ref{spinquasi}), we see that $q_{4}$ is
proportional to a square root coming from $K(1/x)$. We therefore expect that
$\delta q_{4}(x)$ should be proportional to $\partial _{x}K(1/x)\propto
1/K(1/x)$ and make the following ansatz for the fluctuations:
\begin{eqnarray*}
\delta q_{1}(x,y) &=&\frac{\alpha _{+}(y)}{x+1}+\frac{\alpha _{-}(y)}{x-1}%
,\,\,\,\delta q_{2}(x,y)=-\delta q_{1}(1/x,y),\,\,\, \\
\delta q_{5}(x,y) &=&\sum_{\pm }\frac{\alpha (x)}{x-x_{n}^{45}}+\sum_{\pm }%
\frac{\alpha (1/x)}{1/x-x_{n}^{45}}, \\
\delta q_{4}(x,y) &=&h(x,y)/K(1/x),\,\,\,\delta q_{3}(x,y)=-\delta q_{4}(1/x,y).
\end{eqnarray*}%
where $\sum_{\pm }$ stands for the sum over positive and negative mode
number, $y$ is a collective coordinate for $x_{\pm n}^{45}$, $\alpha (x)$ is defined in Eq.~(\ref{eq:alpha}), and $\alpha
_{\pm }(y)$ are some functions to be determined. Note that this ansatz is
consistent with the inversion symmetry in Eq.~(\ref{inversion}). We also
make the following ansatz for $h(x,y)$:
\begin{equation*}
h(x,y)=\frac{\alpha _{+}(y)K(1)}{x+1}+\frac{\alpha _{-}(y)K(1)}{x-1}%
-\sum_{\pm }\frac{\alpha (x_{n}^{45})K(1/x_{n}^{45})}{x-x_{n}^{45}}.
\end{equation*}%
For this choice of $h(x,y)$, the residue of $\delta q_{4}$ at $x=x_{\pm n}^{45}$ agrees with
Eq.~(\ref{eq:alpha}) and the residues of all the fluctuations are
synchronized $x=\pm 1$ according to Eq.~(\ref{eq:synch}). To compute the
anomalous energy shift, we must look at the large-$x$ behavior of the
fluctuations and compare them to Eq.~(\ref{eq:asymptotics}). At large $x$, $%
\delta q_{2}$ and $\delta q_{4}$ are given by
\begin{eqnarray*}
\lim_{x\rightarrow \infty }\delta q_{2}(x,y) &\sim &\alpha _{-}(y)-\alpha
_{+}(y)+\frac{1}{x}\left( \alpha _{+}(y)+\alpha _{-}(y)\right) , \\
\lim_{x\rightarrow \infty }\delta q_{4}(x,y) &\sim &\frac{1}{\mathcal{J}x}%
\left[ K(1)\left( \alpha _{+}(y)+\alpha _{-}(y)\right) -\sum_{\pm }\alpha
(x_{n}^{45})K(1/x_{n}^{45})\right] {\normalcolor.}
\end{eqnarray*}%
where we neglect terms of $\mathcal{O}(x^{-2})$. Comparing the asymptotic
forms of $\delta q_{2}$ and $\delta q_{4}$ with Eq.~(\ref{eq:asymptotics})
gives
\begin{equation*}
\alpha _{+}(y)=\alpha _{-}(y)=\frac{\Delta (y)}{4g}=\frac{1}{\kappa }\left[ -\frac{%
\mathcal{J}}{g}+\sum_{\pm }\alpha \left( x_{n}^{45}\right) K\left(
1/x_{n}^{45}\right) \right] ,
\end{equation*}%
where $\kappa =2K(1)$. Recalling that the (4,5) fluctuation is a $CP^{3}$
fluctuation, Eq.~(\ref{eq:9}) implies that
\begin{equation*}
\Omega _{45}(y)=\frac{1}{K(1)}\left[ \sum_{\pm }\frac{\left(
x_{n}^{45}\right) ^{2}K(1/x_{n}^{45})}{\left( x_{n}^{45}\right) ^{2}-1}%
\right] -\frac{2\mathcal{J}}{K(1)}{\normalcolor.}
\end{equation*}%
This implies that for a single fluctuation
\begin{equation*}
\Omega _{45}(y)=\frac{2}{\kappa }\frac{y^{2}K(1/y)}{y^{2}-1}-\frac{2\mathcal{%
J}}{\kappa }{\normalcolor.}
\end{equation*}%
Now it is trivial to write down all the other off-shell frequencies using
the relations in Eqs.~(\ref{LL},\ref{M35}) and table \ref{tLH}. The off-shell frequencies are summarized in Table~\ref{T5}.
\oddsidemargin -0.7in
\begin{table}[htbp]
\caption{Off-shell frequencies for the fluctuations about the spinning string solution. }\center
\label{T5}
{\small
\begin{equation}
{
\begin{array}{c|c|l}
\toprule  \mathrm{\mathbf{\Omega(y)}} & \mathrm{\mathbf{Pole\,\,\,Location}}
& \mathrm{\mathbf{Polarizations}} \\
\midrule
\begin{array}{l}
\frac{y^{2}+1}{y^{2}-1}%
\end{array}
&
\begin{array}{l}
2\kappa x_{n}=n\left(x_{n}^{2}-1\right)%
\end{array}
&
\begin{array}{l}
\textcolor{red}{\bf{(1,10);(2,9);(1,9)}}%
\end{array}
\\
\midrule
\begin{array}{l}
\frac{1}{2}\frac{y^{2}+1}{y^{2}-1}+\frac{2}{\kappa}\left(\frac{y^{2}K(1/y)}{%
y^{2}-1}-\mathcal{J}\right)  \\
\frac{1}{2}\frac{y^{2}+1}{y^{2}-1}+\frac{2}{\kappa}\frac{K(y)}{y^{2}-1}\\
\frac{1}{2}\frac{y^{2}+1}{y^{2}-1}%
\end{array}
&
\begin{array}{l}
x_{n}\left(\kappa+2K(1/x_{n})\right)=n\left(x_{n}^{2}-1\right) \\
x_{n}\left(\kappa+2K(x_{n})\right)=\left(n+m\right)\left(x_{n}^{2}-1\right)
\\
\kappa x_{n}=n\left(x_{n}^{2}-1\right)
\end{array}
&
\begin{array}{l}
\textcolor{red}{\bf{(1,7);(2,7)}} \\
\textcolor{red}{\bf{(1,8);(2,8)}} \\
\textcolor{blue}{\bf{(1,5);(1,6);(2,5);(2,6)}}\\
\end{array}
\\
\midrule
\begin{array}{l}
\frac{2}{\kappa}\left[\frac{y}{y^{2}-1}\left(K(y)/y+yK(1/y)\right)-\mathcal{J%
}\right] \\
\frac{2}{\kappa}\frac{K(y)}{y^{2}-1} \\
\frac{2}{\kappa}\left(\frac{y^{2}K(1/y)}{y^{2}-1}-\mathcal{J}\right)%
\end{array}
&
\begin{array}{l}
2x_{n}\left(K(x_{n})+K(1/x_{n})\right)=\left(n+m\right)%
\left(x_{n}^{2}-1\right) \\
2x_{n}K(x_{n})=\left(n+m\right)\left(x_{n}^{2}-1\right) \\
2x_{n}K(1/x_{n})=n\left(x_{n}^{2}-1\right)%
\end{array}
&
\begin{array}{l}
\textcolor{red}{\bf{(3,7)}} \\
\textcolor{blue}{\bf{(3,5);(3,6)}} \\
\textcolor{blue}{\bf{(4,5);(4,6)}}%
\end{array}
\\
\bottomrule
\end{array}%
}  \notag
\end{equation}%
}
\end{table}
\oddsidemargin -0.0in

\subsection{On-shell Frequencies}

The structure of this section is as follows: for each row of Table~\ref{T5}, we
find the solutions of the equation in the second column, plug the solution
with greatest magnitude into the off-shell frequency in the first column,
and simplify the resulting expression to obtain the on-shell algebraic curve
frequency in the corresponding row in Table~\ref{tabAC}. We use the following
notation for on-shell frequencies:
\begin{equation*}
\omega_{n}=\Omega\left(x_{n}\right){\normalcolor .}
\end{equation*}

\begin{itemize}
\item \textbf{(1,9); (2,9); (1,10)} \newline
The equation for the pole location implies that $\frac{1}{x_{n}^{2}-1}=\frac{%
n}{2\kappa x_{n}}$. Plugging this into the formula for the off-shell
frequency implies that
\begin{equation}
\omega_{n}=\frac{n}{2\kappa}\left(x_{n}+1/x_{n}\right){\normalcolor .}
\label{a}
\end{equation}
Solving for the pole location gives%
\begin{equation*}
x_{n}=\frac{1}{n}\left(\kappa\pm\sqrt{\kappa^{2}+n^{2}}\right){\normalcolor .%
}
\end{equation*}
Choosing solution with larger magnitude and plugging it into Eq.~(\ref{a})
then leads to
\begin{equation*}
\omega_{n}=\sqrt{1+\frac{n^{2}}{\kappa^{2}}}{\normalcolor .}
\end{equation*}

\item \textbf{(1,7); (2,7)} \newline
The equation for the pole location implies that $\frac{x_{n}K\left(1/x_{n}%
\right)}{x_{n}^{2}-1}=\frac{n}{2}-\frac{\kappa x_{n}}{2\left(x_{n}^{2}-1%
\right)}$. Plugging this into the off-shell frequency and doing a little
algebra gives%
\begin{equation}
\omega_{n}=\frac{n}{\kappa}x_{n}-\frac{2\mathcal{J}}{\kappa}-1/2.
\label{eq:17}
\end{equation}
Solving for the pole location gives%
\begin{equation*}
x_{n}=\frac{1}{n}\left(\kappa\pm\sqrt{4\mathcal{J}^{2}+n^{2}}\right){%
\normalcolor .}
\end{equation*}
Taking the solution with larger magnitude and plugging it onto Eq.~(\ref%
{eq:17}) then gives
\begin{equation*}
\frac{1}{\kappa}\left(\sqrt{4\mathcal{J}^{2}+n^{2}}-2\mathcal{J}\right)+%
\frac{1}{2}{\normalcolor .}
\end{equation*}

\item \textbf{(1,8); (2,8)} \newline
From the equation for the pole location we find $\frac{K(x_{n})}{x_{n}^{2}-1}%
=\frac{1}{2x_{n}}\left(n+m\right)-\frac{\kappa}{2\left(x_{n}^{2}-1%
\right)}$. Plugging this into the off-shell frequency and doing a little
algebra gives%
\begin{equation}
\omega_{n}=\frac{n+m}{\kappa x_{n}}+1/2.  \label{eq:18}
\end{equation}
The solutions to the equation for the pole location are
\begin{equation*}
x_{n}=\frac{(m+n)}{n(2m+n)}\left(\kappa\pm\sqrt{4\mathcal{J}^{2}+(m+n)^{2}}%
\right){\normalcolor .}
\end{equation*}
Taking the solution with larger magnitude and plugging it into Eq.~(\ref%
{eq:18}) gives%
\begin{equation*}
\omega_{n}=\frac{n(2m+n)}{\kappa}\frac{1}{\kappa+\sqrt{4\mathcal{J}%
^{2}+(m+n)^{2}}}+1/2{\normalcolor .}
\end{equation*}
Finally, multiplying the numerator and denominator in first term by $\kappa-%
\sqrt{4\mathcal{J}^{2}+(m+n)^{2}}$ and doing a little more algebra gives
\begin{equation*}
\omega_{n}=\frac{1}{\kappa}\sqrt{4\mathcal{J}^{2}+\left(m+n\right)^{2}}-%
\frac{1}{2}{\normalcolor .}
\end{equation*}

\item \textbf{(1,5); (1,6); (2,5); (2,6)}\newline
This is very similar to the calculation for 19, 29, 1 10, so we omit it.

\item \textbf{(3,7)} \newline
From the equation for the pole location, we have $\frac{2x_{n}}{x_{n}^{2}-1}=%
\frac{n+m}{K(x_{n})+K(1/x_{n})}$. Plugging this into the off-shell
frequency gives
\begin{equation}
\omega_{n}=\frac{1}{\kappa}\left[\left(n+m\right)\beta-2\mathcal{J}%
\right],\,\,\,\beta=\frac{\frac{1}{x_{n}}K(x_{n})+x_{n}K(1/x_{n})}{%
K(x_{n})+K(1/x_{n})}{\normalcolor .}  \label{b}
\end{equation}
Let's focus on the term $\beta$. Multiplying the numerator and denominator
by $K(x_{n})-K(1/x_{n})$ gives
\begin{equation}
\beta=\frac{\frac{1}{x_{n}}%
K(x_{n})^{2}-x_{n}K(1/x_{n})^{2}+(x_{n}-1/x_{n})K(x_{n})K(1/x_{n})}{%
m^{2}(x_{n}-1/x_{n})(x_{n}+1/x_{n})/4}{\normalcolor .}  \label{c}
\end{equation}
By squaring the equation for the pole location, we find that
\begin{equation*}
K(x_{n})K(1/x_{n})=\frac{1}{2}\left[\frac{1}{4}\left(n+m%
\right)^{2}(x_{n}-1/x_{n})^{2}-K(x_{n})^{2}-K(1/x_{n})^{2}\right]{%
\normalcolor .}
\end{equation*}
Plugging this into Eq.~(\ref{c}) and doing some algebra gives%
\begin{equation*}
\beta=\frac{\frac{1}{2}m^{2}\left[3(x_{n}-1/x_{n})+1/x_{n}^{3}-x_{n}^{3}%
\right]+8\mathcal{J}^{2}(1/x_{n}-x_{n})+\frac{1}{2}\left(n%
+m\right)^{2}(x_{n}-1/x_{n})^{3}}{m^{2}(x_{n}-1/x_{n})(x_{n}+1/x_{n})}{%
\normalcolor .}
\end{equation*}
Noting that $x^{3}-1/x^{3}=\left(x^{2}+1/x^{2}+1\right)(x-1/x)$ and doing
some more algebra then gives
\begin{equation*}
\beta=\frac{-8\mathcal{J}^{2}+n\left(\frac{n}{2}%
+m\right)\left(x_{n}-1/x_{n}\right)^{2}}{m^{2}(x_{n}+1/x_{n})}{\normalcolor .%
}
\end{equation*}
Noting that $(x-1/x)^{2}=(x+1/x)^{2}-4$ finally gives
\begin{equation*}
\beta=\frac{n\left(n/2+m\right)}{m^{2}}\left(x+1/x\right)-\frac{%
4n\left(n/2+m\right)+8\mathcal{J}^{2}}{m^{2}\left(x+1/x\right)}{\normalcolor %
.}
\end{equation*}
Combining this with Eq.~(\ref{b}), we find
\begin{equation}
\omega _{n}=\frac{1}{\kappa }\left[ \frac{n+m}{m^{2}}\left[
\begin{array}{c}
n(m+n/2)(x_{n}+1/x_{n}) \\
-\left( 8\mathcal{J}^{2}+2n(2m+n)\right) (x_{n}+1/x_{n})^{-1}%
\end{array}%
\right] -2\mathcal{J}\right] {\normalcolor.}  \label{d}
\end{equation}%
The solutions for the pole location are
\begin{equation*}
x_{n}=\pm \frac{1}{n(2m+n)}\left[
\begin{array}{c}
8\mathcal{J}^{2}(m+n)^{2}+n(2m+n)\left( 2m^{2}+n(2m+n)\right)  \\
\pm 4|m+n|\sqrt{\left( 4\mathcal{J}^{2}+n(2m+n)\right) \left( \mathcal{J}%
^{2}(m+n)^{2}+m^{2}n(2m+n)/4\right) }%
\end{array}%
\right] ^{1/2}{\normalcolor.}
\end{equation*}
Taking the solution with $+$ sign out front, we see that for either choice
of sign inside square root we have%
\begin{equation*}
x_{n}+1/x_{n}=\frac{2(m+n)}{n(2m+n)}\sqrt{4\mathcal{J}^{2}+n(2m+n)}{%
\normalcolor .}
\end{equation*}
Plugging this into Eq.~(\ref{d}) and doing a little more algebra finally
gives%
\begin{equation*}
\omega_{n}=\frac{1}{\kappa}\left(\sqrt{4\mathcal{J}^{2}-m^{2}+(m+n)^{2}}-2%
\mathcal{J}\right){\normalcolor .}
\end{equation*}

\item \textbf{(3,5); (3,6)} \newline
The equation for the pole location implies that $\frac{K\left(x_{n}\right)}{%
x_{n}^{2}-1}=\left(n+m\right)\frac{1}{2x_{n}}$. Using this in the
formula for the off-shell frequency leads to
\begin{equation}
\omega_{n}=\frac{1}{\kappa}\left(n+m\right)\frac{1}{x_{n}}{%
\normalcolor .}  \label{eq:35}
\end{equation}
Solving the equation for the pole location gives%
\begin{equation*}
x_{n}=\pm\frac{m+n}{\sqrt{2\mathcal{J}^{2}+\left(m+n\right)^{2}\pm\sqrt{4%
\mathcal{J}^{4}+4\kappa^{2}\left(m+n\right)^{2}}}}{\normalcolor .}
\end{equation*}
The solution with larger magnitude is the one with relative $-$ sign in
denominator. Taking the solution with greater magnitude and $+$ sign out
front and plugging this into Eq.~(\ref{eq:35}) gives
\begin{equation*}
\omega_{n}=\frac{1}{\kappa}\sqrt{2\mathcal{J}^{2}+\left(m+n\right)^{2}-%
\sqrt{4\mathcal{J}^{4}+\left(m+n\right)^{2}\kappa^{2}}}{\normalcolor .}
\end{equation*}

\item \textbf{(4,5); (4,6)} \newline
Plugging the equation for the pole location into the off-shell frequency
gives%
\begin{equation}
\omega_{n}=\frac{n}{\kappa}x_{n}-\frac{2\mathcal{J}}{\kappa}{\normalcolor .}
\label{eq:45}
\end{equation}
The solutions for the pole location are%
\begin{equation*}
x_{n}=\pm\frac{1}{n}\sqrt{2\mathcal{J}^{2}+n^{2}\pm\sqrt{4\mathcal{J}%
^{4}+n^{2}\kappa^{2}}}{\normalcolor .}
\end{equation*}
If we choose the solution with greater magnitude and plug it into Eq.~(\ref%
{eq:45}), we have
\begin{equation*}
\frac{1}{\kappa}\left(\sqrt{2\mathcal{J}^{2}+n^{2}+\sqrt{4\mathcal{J}%
^{4}+n^{2}\kappa^{2}}}-2\mathcal{J}\right){\normalcolor .}
\end{equation*}
\end{itemize}

\section{Comparison with Bethe Ansatz}

In this Appendix we will compute the leading two contributions to the anomalous dimension of the operator dual to the spinning string in $AdS_{4}\times CP^{3}$. First let's consider the operator dual to the $SU(2)$ spinning string in $AdS_{5}\times S^{5}$ which has the form
\begin{equation}
\mathcal{O}={\rm tr}\left[Z^{J}W^{J}+\text{permutations}\right],
\label{su2op}
\end{equation}
where $Z$ and $W$ are complex scalar fields in $\mathcal{N}=4$ SYM. In this sector, the one-loop planar dilatation operator corresponds to the Hamiltonian of a Heisenberg spin chain of length $2J$ \cite{Minahan:2002ve}:
\begin{equation}
\mathbf{\Delta }-2J=\frac{\lambda }{8\pi ^{2}}\sum_{i=1}^{2J}\left(
1-P_{i,i+1}\right) {\normalcolor.}  \label{eq:s5su2}
\end{equation}%
The dilatation operator can be diagonalized by solving a set of Bethe ansatz equations \cite{Bethe}:
\begin{subequations}
\label{BA}
\begin{eqnarray}
\left( \frac{u_{j}+i/2}{u_{j}-i/2}\right) ^{2J} &=&\prod_{k\neq j}^{J}\frac{u_{j}-u_{k}+i}{u_{j}-u_{k}-i}, \\
\prod_{j=1}^{J}\left( \frac{u_{j}+i/2}{u_{j}-i/2}\right)
&=&1\Longrightarrow \sum_{j=1}^{J}\ln \left( \frac{u_{j}+i/2}{u_{j}-i/2}%
\right) =-2\pi mi, \label{BA1} \\
\Delta -2J &=&\frac{\lambda }{8\pi ^{2}}\sum_{j=1}^{J}\frac{1}{%
u_{j}^{2}+1/4}
\end{eqnarray}
where $m$ is an integer which is introduced after taking the log of both sides of equation \ref{BA1}.
In the large-$J$ limit, the Bethe equations simplify and can be solved using the methods described in \cite{Kazakov:2004qf,Lubcke:2004dg,Beisert:2005mq}. In particular, \cite{Beisert:2005mq} found that the anomalous dimension is given by
\end{subequations}
\begin{eqnarray}
\Delta -2J &=&\left( \frac{\lambda m^{2}}{4J}+\ldots \right) +\frac{1}{J}%
\left( \frac{a\lambda }{8J}+\ldots \right) ,  \label{AD} \\
a &=&m^{2}+\sum_{n=1}^{\infty }\left( n\sqrt{n^{2}-4m^{2}}%
-n^{2}+2m^{2}\right) .  \nonumber
\end{eqnarray}%

Now let's turn to the operator in Eq. (\ref{dualop}). In this case, the two-loop planar dilatation operator is given by Eq. (\ref{eq:cp3su2}). As explained in section 4.1, this corresponds to the Hamiltonian for two identical Heisenberg spin
chains of length $2J$ which are only coupled by a momentum constraint. With this in mind, the Bethe equations are
\begin{subequations}
\label{BA2}
\begin{eqnarray}
\left( \frac{u_{j}+i/2}{u_{j}-i/2}\right) ^{2J} &=&\prod_{k\neq j}^{J}\frac{%
u_{j}-u_{k}+i}{u_{j}-u_{k}-i}, \\
\left( \prod_{j=1}^{J}\left( \frac{u_{j}+i/2}{u_{j}-i/2}\right) \right)
^{2} &=&1\Longrightarrow \sum_{j=1}^{J}\ln \left( \frac{u_{j}+i/2}{u_{j}-i/2%
}\right) =-\pi mi, \\
\Delta -2J &=&2\lambda ^{2}\sum_{j=1}^{J}\frac{1}{u_{j}^{2}+1/4}.
\end{eqnarray}%
Comparing both sets of Bethe equations we see
that Eq. (\ref{BA}) can be mapped into Eq.(\ref{BA2}) by making the following relabeling:
\end{subequations}
\[\qquad m\rightarrow m/2,\qquad \lambda \rightarrow 16\pi^{2}\lambda ^{2}.
\]%
Making these substitutions in Eq.(\ref{AD}) gives Eq.(\ref{prediction}), which we obtained using string theory.


\begin{thebibliography}{10}

\bibitem{Maldacena:1997re}
J.~M. Maldacena, {\it {The large N limit of superconformal field theories and
  supergravity}},  {\em Adv. Theor. Math. Phys.} {\bf 2} (1998) 231--252,
  [\href{http://xxx.lanl.gov/abs/hep-th/9711200}{{\tt hep-th/9711200}}].

\bibitem{Aharony:2008ug}
O.~Aharony, O.~Bergman, D.~L. Jafferis, and J.~Maldacena, {\it {N=6
  superconformal Chern-Simons-matter theories, M2-branes and their gravity
  duals}},  {\em JHEP} {\bf 10} (2008) 091,
  [\href{http://xxx.lanl.gov/abs/0806.1218}{{\tt arXiv:0806.1218}}].

\bibitem{Minahan:2008hf}
J.~A. Minahan and K.~Zarembo, {\it {The Bethe ansatz for superconformal
  Chern-Simons}},  {\em JHEP} {\bf 09} (2008) 040,
  [\href{http://xxx.lanl.gov/abs/0806.3951}{{\tt arXiv:0806.3951}}].

\bibitem{Zwiebel:2009vb}
B.~I. Zwiebel, {\it {Two-loop Integrability of Planar N=6 Superconformal Chern-
  Simons Theory}},  {\em J. Phys.} {\bf A42} (2009) 495402,
  [\href{http://xxx.lanl.gov/abs/0901.0411}{{\tt arXiv:0901.0411}}].

\bibitem{Minahan:2009te}
J.~A. Minahan, W.~Schulgin, and K.~Zarembo, {\it {Two loop integrability for
  Chern-Simons theories with N=6 supersymmetry}},  {\em JHEP} {\bf 03} (2009)
  057, [\href{http://xxx.lanl.gov/abs/0901.1142}{{\tt arXiv:0901.1142}}].

\bibitem{Bak:2009mq}
D.~Bak, H.~Min, and S.-J. Rey, {\it {Generalized Dynamical Spin Chain and
  4-Loop Integrability in N=6 Superconformal Chern-Simons Theory}},  {\em Nucl.
  Phys.} {\bf B827} (2010) 381--405,
  [\href{http://xxx.lanl.gov/abs/0904.4677}{{\tt arXiv:0904.4677}}].

\bibitem{Bena:2003wd}
I.~Bena, J.~Polchinski, and R.~Roiban, {\it {Hidden symmetries of the AdS(5) x
  S**5 superstring}},  {\em Phys. Rev.} {\bf D69} (2004) 046002,
  [\href{http://xxx.lanl.gov/abs/hep-th/0305116}{{\tt hep-th/0305116}}].

\bibitem{Stefanski:2008ik}
B.~Stefanski, jr, {\it {Green-Schwarz action for Type IIA strings on
  $AdS_4\times CP^3$}},  {\em Nucl. Phys.} {\bf B808} (2009) 80--87,
  [\href{http://xxx.lanl.gov/abs/0806.4948}{{\tt arXiv:0806.4948}}].

\bibitem{Arutyunov:2008if}
G.~Arutyunov and S.~Frolov, {\it {Superstrings on $AdS_{4} \times CP^{3}$ as a
  Coset Sigma-model}},  {\em JHEP} {\bf 09} (2008) 129,
  [\href{http://xxx.lanl.gov/abs/0806.4940}{{\tt arXiv:0806.4940}}].

\bibitem{D'Auria:2008cw}
R.~D'Auria, P.~Fre, P.~A. Grassi, and M.~Trigiante, {\it {Superstrings on
  $AdS_4 x CP^3$ from Supergravity}},  {\em Phys. Rev.} {\bf D79} (2009)
  086001, [\href{http://xxx.lanl.gov/abs/0808.1282}{{\tt arXiv:0808.1282}}].

\bibitem{Gomis:2008jt}
J.~Gomis, D.~Sorokin, and L.~Wulff, {\it {The complete AdS(4) x CP(3)
  superspace for the type IIA superstring and D-branes}},  {\em JHEP} {\bf 03}
  (2009) 015, [\href{http://xxx.lanl.gov/abs/0811.1566}{{\tt
  arXiv:0811.1566}}].

\bibitem{Fre:2008qc}
P.~Fre and P.~A. Grassi, {\it {Pure Spinor Formalism for {Osp}(N|4)
  backgrounds}},  \href{http://xxx.lanl.gov/abs/0807.0044}{{\tt
  arXiv:0807.0044}}.

\bibitem{Bonelli:2008us}
G.~Bonelli, P.~A. Grassi, and H.~Safaai, {\it {Exploring Pure Spinor String
  Theory on $AdS_4\times \mathbb{CP}^3$}},  {\em JHEP} {\bf 10} (2008) 085,
  [\href{http://xxx.lanl.gov/abs/0808.1051}{{\tt arXiv:0808.1051}}].

\bibitem{SchaferNameki:2004ik}
S.~Schafer-Nameki, {\it {The algebraic curve of 1-loop planar N = 4 SYM}},
  {\em Nucl. Phys.} {\bf B714} (2005) 3--29,
  [\href{http://xxx.lanl.gov/abs/hep-th/0412254}{{\tt hep-th/0412254}}].

\bibitem{Kazakov:2004qf}
V.~A. Kazakov, A.~Marshakov, J.~A. Minahan, and K.~Zarembo, {\it {Classical /
  quantum integrability in AdS/CFT}},  {\em JHEP} {\bf 05} (2004) 024,
  [\href{http://xxx.lanl.gov/abs/hep-th/0402207}{{\tt hep-th/0402207}}].

\bibitem{Beisert:2005bm}
N.~Beisert, V.~A. Kazakov, K.~Sakai, and K.~Zarembo, {\it {The algebraic curve
  of classical superstrings on AdS(5) x S**5}},  {\em Commun. Math. Phys.} {\bf
  263} (2006) 659--710, [\href{http://xxx.lanl.gov/abs/hep-th/0502226}{{\tt
  hep-th/0502226}}].

\bibitem{Gromov:2008bz}
N.~Gromov and P.~Vieira, {\it {The AdS4/CFT3 algebraic curve}},  {\em JHEP}
  {\bf 02} (2009) 040, [\href{http://xxx.lanl.gov/abs/0807.0437}{{\tt
  arXiv:0807.0437}}].

\bibitem{AllLoop}
N.~Gromov and P.~Vieira, {\it The all loop ads4/cft3 bethe ansatz},  {\em
  Journal of High Energy Physics} {\bf 2009} (2009), no.~01 016.

\bibitem{Nishioka:2008gz}
T.~Nishioka and T.~Takayanagi, {\it {On Type IIA Penrose Limit and N=6
  Chern-Simons Theories}},  {\em JHEP} {\bf 08} (2008) 001,
  [\href{http://xxx.lanl.gov/abs/0806.3391}{{\tt arXiv:0806.3391}}].

\bibitem{Grignani:2008is}
G.~Grignani, T.~Harmark, and M.~Orselli, {\it {The SU(2) x SU(2) sector in the
  string dual of N=6 superconformal Chern-Simons theory}},  {\em Nucl. Phys.}
  {\bf B810} (2009) 115--134, [\href{http://xxx.lanl.gov/abs/0806.4959}{{\tt
  arXiv:0806.4959}}].

\bibitem{Gaiotto:2008cg}
D.~Gaiotto, S.~Giombi, and X.~Yin, {\it {Spin Chains in N=6 Superconformal
  Chern-Simons-Matter Theory}},  {\em JHEP} {\bf 04} (2009) 066,
  [\href{http://xxx.lanl.gov/abs/0806.4589}{{\tt arXiv:0806.4589}}].

\bibitem{Berenstein:2002jq}
D.~E. Berenstein, J.~M. Maldacena, and H.~S. Nastase, {\it {Strings in flat
  space and pp waves from N = 4 super Yang Mills}},  {\em JHEP} {\bf 04} (2002)
  013, [\href{http://xxx.lanl.gov/abs/hep-th/0202021}{{\tt hep-th/0202021}}].

\bibitem{Ahn:2008aa}
C.~Ahn and R.~I. Nepomechie, {\it {N=6 super Chern-Simons theory S-matrix and
  all-loop Bethe ansatz equations}},  {\em JHEP} {\bf 09} (2008) 010,
  [\href{http://xxx.lanl.gov/abs/0807.1924}{{\tt arXiv:0807.1924}}].

\bibitem{Zarembo:2009au}
K.~Zarembo, {\it {Worldsheet spectrum in AdS(4)/CFT(3) correspondence}},
  \href{http://xxx.lanl.gov/abs/0903.1747}{{\tt arXiv:0903.1747}}.

\bibitem{Sundin:2009zu}
P.~Sundin, {\it {On the worldsheet theory of the type IIA AdS(4) x CP(3)
  superstring}},  \href{http://xxx.lanl.gov/abs/0909.0697}{{\tt
  arXiv:0909.0697}}.

\bibitem{Bergman:2009zh}
O.~Bergman and S.~Hirano, {\it {Anomalous radius shift in AdS(4)/CFT(3)}},
  {\em JHEP} {\bf 07} (2009) 016,
  [\href{http://xxx.lanl.gov/abs/0902.1743}{{\tt arXiv:0902.1743}}].

\bibitem{Krishnan:2008zs}
C.~Krishnan, {\it {AdS4/CFT3 at One Loop}},  {\em JHEP} {\bf 09} (2008) 092,
  [\href{http://xxx.lanl.gov/abs/0807.4561}{{\tt arXiv:0807.4561}}].

\bibitem{McLoughlin:2008ms}
T.~McLoughlin and R.~Roiban, {\it {Spinning strings at one-loop in $AdS_{4}
  \times P^3$}},  {\em JHEP} {\bf 12} (2008) 101,
  [\href{http://xxx.lanl.gov/abs/0807.3965}{{\tt arXiv:0807.3965}}].

\bibitem{Alday:2008ut}
L.~F. Alday, G.~Arutyunov, and D.~Bykov, {\it {Semiclassical Quantization of
  Spinning Strings in $AdS_{4} \times CP^{3}$}},  {\em JHEP} {\bf 11} (2008)
  089, [\href{http://xxx.lanl.gov/abs/0807.4400}{{\tt arXiv:0807.4400}}].

\bibitem{Gromov:2008fy}
N.~Gromov and V.~Mikhaylov, {\it {Comment on the Scaling Function in AdS4 x
  CP3}},  {\em JHEP} {\bf 04} (2009) 083,
  [\href{http://xxx.lanl.gov/abs/0807.4897}{{\tt arXiv:0807.4897}}].

\bibitem{McLoughlin:2008he}
T.~McLoughlin, R.~Roiban, and A.~A. Tseytlin, {\it {Quantum spinning strings in
  $AdS_4 x CP^3$: testing the Bethe Ansatz proposal}},  {\em JHEP} {\bf 11}
  (2008) 069, [\href{http://xxx.lanl.gov/abs/0809.4038}{{\tt
  arXiv:0809.4038}}].

\bibitem{Shenderovich:2008bs}
I.~Shenderovich, {\it {Giant magnons in $AdS_4/CFT_3$: dispersion, quantization
  and finite--size corrections}},
  \href{http://xxx.lanl.gov/abs/0807.2861}{{\tt arXiv:0807.2861}}.

\bibitem{Frolov:2003qc}
S.~Frolov and A.~A. Tseytlin, {\it {Multi-spin string solutions in AdS(5) x
  S**5}},  {\em Nucl. Phys.} {\bf B668} (2003) 77--110,
  [\href{http://xxx.lanl.gov/abs/hep-th/0304255}{{\tt hep-th/0304255}}].

\bibitem{Frolov:2003tu}
S.~Frolov and A.~A. Tseytlin, {\it {Quantizing three-spin string solution in
  AdS(5) x S**5}},  {\em JHEP} {\bf 07} (2003) 016,
  [\href{http://xxx.lanl.gov/abs/hep-th/0306130}{{\tt hep-th/0306130}}].

\bibitem{Frolov:2004bh}
S.~A. Frolov, I.~Y. Park, and A.~A. Tseytlin, {\it {On one-loop correction to
  energy of spinning strings in S(5)}},  {\em Phys. Rev.} {\bf D71} (2005)
  026006, [\href{http://xxx.lanl.gov/abs/hep-th/0408187}{{\tt
  hep-th/0408187}}].

\bibitem{Beisert:2005mq}
N.~Beisert, A.~A. Tseytlin, and K.~Zarembo, {\it {Matching quantum strings to
  quantum spins: One-loop vs. finite-size corrections}},  {\em Nucl. Phys.}
  {\bf B715} (2005) 190--210,
  [\href{http://xxx.lanl.gov/abs/hep-th/0502173}{{\tt hep-th/0502173}}].

\bibitem{Chen:2008qq}
B.~Chen and J.-B. Wu, {\it {Semi-classical strings in $AdS_{4} \times
  CP^{3}$}},  {\em JHEP} {\bf 09} (2008) 096,
  [\href{http://xxx.lanl.gov/abs/0807.0802}{{\tt arXiv:0807.0802}}].

\bibitem{Ahn:2008hj}
C.~Ahn, P.~Bozhilov, and R.~C. Rashkov, {\it {Neumann-Rosochatius integrable
  system for strings on $AdS_4 x CP^3$}},  {\em JHEP} {\bf 09} (2008) 017,
  [\href{http://xxx.lanl.gov/abs/0807.3134}{{\tt arXiv:0807.3134}}].

\bibitem{Gromov:2007aq}
N.~Gromov and P.~Vieira, {\it {The AdS(5) x S**5 superstring quantum spectrum
  from the algebraic curve}},  {\em Nucl. Phys.} {\bf B789} (2008) 175--208,
  [\href{http://xxx.lanl.gov/abs/hep-th/0703191}{{\tt hep-th/0703191}}].

\bibitem{Frolov:2002av}
S.~Frolov and A.~A. Tseytlin, {\it {Semiclassical quantization of rotating
  superstring in AdS(5) x S(5)}},  {\em JHEP} {\bf 06} (2002) 007,
  [\href{http://xxx.lanl.gov/abs/hep-th/0204226}{{\tt hep-th/0204226}}].

\bibitem{Cvetic:1999zs}
M.~Cvetic, H.~Lu, C.~N. Pope, and K.~S. Stelle, {\it {T-Duality in the
  Green-Schwarz Formalism, and the Massless/Massive IIA Duality Map}},  {\em
  Nucl. Phys.} {\bf B573} (2000) 149--176,
  [\href{http://xxx.lanl.gov/abs/hep-th/9907202}{{\tt hep-th/9907202}}].

\bibitem{Gromov:2008ec}
N.~Gromov, S.~Schafer-Nameki, and P.~Vieira, {\it {Efficient precision
  quantization in AdS/CFT}},  {\em JHEP} {\bf 12} (2008) 013,
  [\href{http://xxx.lanl.gov/abs/0807.4752}{{\tt arXiv:0807.4752}}].

\bibitem{notes}
V.~Mikhaylov, {\it unpublished}, .

\bibitem{SchaferNameki:2005tn}
S.~Schafer-Nameki, M.~Zamaklar, and K.~Zarembo, {\it {Quantum corrections to
  spinning strings in AdS(5) x S**5 and Bethe ansatz: A comparative study}},
  {\em JHEP} {\bf 09} (2005) 051,
  [\href{http://xxx.lanl.gov/abs/hep-th/0507189}{{\tt hep-th/0507189}}].

\bibitem{Tseytlin:2003ii}
A.~A. Tseytlin, {\it {Spinning strings and AdS/CFT duality}},
  \href{http://xxx.lanl.gov/abs/hep-th/0311139}{{\tt hep-th/0311139}}.

\bibitem{SchaferNameki:2006gk}
S.~Schafer-Nameki, {\it {Exact expressions for quantum corrections to spinning
  strings}},  {\em Phys. Lett.} {\bf B639} (2006) 571--578,
  [\href{http://xxx.lanl.gov/abs/hep-th/0602214}{{\tt hep-th/0602214}}].

\bibitem{Metsaev:2002re}
R.~R. Metsaev and A.~A. Tseytlin, {\it {Exactly solvable model of superstring
  in plane wave Ramond-Ramond background}},  {\em Phys. Rev.} {\bf D65} (2002)
  126004, [\href{http://xxx.lanl.gov/abs/hep-th/0202109}{{\tt
  hep-th/0202109}}].

\bibitem{Bandres:2008ry}
M.~A. Bandres, A.~E. Lipstein, and J.~H. Schwarz, {\it {Studies of the ABJM
  Theory in a Formulation with Manifest SU(4) R-Symmetry}},  {\em JHEP} {\bf
  09} (2008) 027, [\href{http://xxx.lanl.gov/abs/0807.0880}{{\tt
  arXiv:0807.0880}}].

\bibitem{Benna:2008zy}
M.~Benna, I.~Klebanov, T.~Klose, and M.~Smedback, {\it {Superconformal
  Chern-Simons Theories and $AdS_{4}/CFT_{3}$ Correspondence}},  {\em JHEP}
  {\bf 09} (2008) 072, [\href{http://xxx.lanl.gov/abs/0806.1519}{{\tt
  arXiv:0806.1519}}].

\bibitem{Cvetic:2000yp}
M.~Cvetic, H.~Lu, and C.~N. Pope, {\it {Consistent warped-space Kaluza-Klein
  reductions, half- maximal gauged supergravities and CP(n) constructions}},
  {\em Nucl. Phys.} {\bf B597} (2001) 172--196,
  [\href{http://xxx.lanl.gov/abs/hep-th/0007109}{{\tt hep-th/0007109}}].

\bibitem{Minahan:2002ve}
J.~A. Minahan and K.~Zarembo, {\it {The Bethe-ansatz for N = 4 super
  Yang-Mills}},  {\em JHEP} {\bf 03} (2003) 013,
  [\href{http://xxx.lanl.gov/abs/hep-th/0212208}{{\tt hep-th/0212208}}].

\bibitem{Bethe}
H. Bethe, {\it On the Theory of Metals. 1. Eigenvalues and Eigenfunctions for the Linear Atomic Chain}, Z. Phys. {\bf 71}, 205 (1931).

\bibitem{Lubcke:2004dg}
M.~Lubcke and K.~Zarembo, {\it {Finite-size corrections to anomalous dimensions
  in N = 4 SYM theory}},  {\em JHEP} {\bf 05} (2004) 049,
  [\href{http://xxx.lanl.gov/abs/hep-th/0405055}{{\tt hep-th/0405055}}].

\end{thebibliography}

\providecommand{\href}[2]{#2}\begingroup\raggedright\endgroup

\end{document}